\RequirePackage{booktabs} 
\documentclass[pdflatex,sn-apa,a4paper]{sn-jnl}
\usepackage[T1]{fontenc}
\usepackage{amsmath,amsfonts}
\usepackage{algorithmic}
\usepackage{array}
\usepackage{makecell}
\usepackage{hhline}
\usepackage[caption=false,font=normalsize,labelfont=sf,textfont=sf]{subfig}
\usepackage{textcomp}
\usepackage{stfloats}
\usepackage{url}
\usepackage{verbatim}
\usepackage{graphicx}
\usepackage{multirow,subfig}

\title{Solving rescheduling problems in heterogeneous urban railway networks using hybrid quantum-classical approach}

\author[1]{M\'aty\'as Koniorczyk \email{koniorczyk.matyas@wigner.hun-ren.hu}}

\author[2]{Krzysztof Krawiec \email{krzysztof.krawiec@polsl.pl}}

\author[3,4]{Ludmila Botelho \email{lbotelho@iitis.pl}}

\author[5]{Nikola Be\v{s}inovi\'c \email{nikola.besinovic@tu-dresden.de}}

\author[3]{Krzysztof Domino \email{kdomino@iitis.pl}}

\affil[1]{HUN-REN Wigner Research Centre for Physics, Konkoly-Thege Mikl\'os \'ut 29-33., Budapest, 1121, 
	Hungary}

\affil[2]{Faculty of Transport and Aviation Engineering, Silesian University of Technology, Akademicka 2A, Gliwice, 44-100, Poland}

\affil[3]{Institute of Theoretical and Applied Informatics, Polish Academy of Sciences, Ba{\l}tycka 5, Gliwice, 44-100, Poland}

\affil[4]{Joint Doctoral School, Silesian University of Technology, Akademicka 2A, Gliwice, 44-100, Poland}

\affil[5]{``Friedrich List'' Faculty of Transport and Traffic Sciences, Technical University of Dresden, 
	Dresden, 01069, Germany}

\begin{document}

	\maketitle

\begin{abstract}

We address the applicability of a hybrid quantum-classical heuristics for practical railway rescheduling management problems. 
We build an integer linear programming model and solve it with D-Wave's quantum-classical hybrid solver (CQM) as well as with CPLEX, for comparison.
The proposed approach is demonstrated on a real-life heterogeneous urban network in Poland, including both single- and multi-track segments. All the requirements posed by the operator of the network are included in the model.
The computational results demonstrate the readiness for application and the benefits of quantum-classical hybrid solvers in a realistic railway scenario: they yield acceptable solutions on time, which is a critical requirement in a rescheduling situation. In particular, CQM as a probabilistic heuristic solver provides a number of feasible, close-to-optimal solutions the dispatcher can choose from.
\end{abstract}

\section*{Keywords}
	railway rescheduling \sep conflict management \sep heterogeneous urban railway network \sep quantum annealing\sep
	hybrid quantum-classical heuristics

\section{Introduction}

Rail transport is expected to experience an increase in capacity demands due to changes in mobility needs resulting from climate policies, leading to traffic challenges in passenger and cargo rail transport. The situation is aggravated by the fact that the rolling stock and, especially, railway infrastructure cannot keep up with the increase in transport needs, which overloads railway systems.
Rail transport, due to its technical and organizational characteristics, is very sensitive to disturbances and disruptions in traffic. These extraordinary events have an impact on railway operations, typically resulting in delays \citep{Ge2022}. Examples of such disturbances include late train departures/arrivals, extended dwell times, etc. Their duration can be from several minutes up to hours.
A \emph{disturbance} is a smaller perturbation that can be handled solely by modifying the existing train paths, whereas a disruption affects rolling stock and crew schedules, too~\citep{cacchiani2014overview}.
In the present work, we focus on train rescheduling from the network operator's perspective. 
Even though we also consider (partial) track closures, which are typically viewed as disruptions, in our examples they will be managed by modifying existing train paths.
The impact of disturbances and disruptions can propagate to multiple sections in the railway network~(c.f. \cite{tornquist2007railway} and references therein). 
Ensuring stable railway traffic and providing reliable service for passengers, rail cargo companies, and their clients is in the best common interest of railway infrastructure managers and train operators. 

A railway network is a complex non-local structure.
Hence, modeling a
bigger portion of it is necessary for the efficient suppression of the
consequences caused by disturbances. 
Diverse objective
functions have been employed for that purpose, such as the (weighted) sum of delays
\citep{lange2018approaches}, the maximal delay that cannot be
avoided~\citep{DarianoPDPbb}, or fuel consumption measures
~\citep{harrod2011modeling}.  
These large-scale rescheduling problems have to be solved almost in real time, as the time available for making decisions is limited.
The railway dispatching/rescheduling problem is recognized as
equivalent to job-shop scheduling with blocking and no-wait
constraints~\citep{Szpigel1973,MASCIS2002498}.  This scheduling problem has an extensive
literature, already summarized in a number of reviews, including those
by~\cite{cacchiani2014overview}, by~\cite{Lamorgese2018}, or
by~\cite{10079880}.

One of the most frequently applied modeling strategies is based on
alternative graphs~\citep{DarianoPDPbb, MASCIS2002498}, resulting in a disjunctive program. Such a model can also be converted into a mixed integer linear program referred to as
the big-M model. As this model often becomes too big to be solved on
time, a number of strategies are applied in order to quickly find a suitable solution, including heuristics,
specific branch-and-bound methods, alternative formulations,
decompositions, and combinations of these; see the work
of~\cite{Lamorgese2018} for a review of the model and the methods.

An alternative approach to this scheduling problem is to use time-indexed models: discrete-time
units and binary decision variables that assign events to particular
time instants. Even though the so-arising $0-1$ programs can be large,
this method has been applied for both timetabling and
dispatching/rescheduling~\citep{CAIMI20122578,LUSBY2013713,MENG2014208}.
Meanwhile, $0-1$ programs are suitable as input
for certain quantum hardware types. Indeed, job-shop problems
addressed on hardware quantum annelaers~\citep{venturelli2016job}, and
even in certain hybrid approaches~\citep{10.1007/978-3-030-50433-5_39}
use similar models. Our previous approach~\citep{domino2023quantum}, which was the first to apply a quantum solver to railway dispatching problems, also followed this route.  In the present work, however, we
deal with a variant of the big-M model, albeit with discrete time
variables, resulting in an integer linear program (ILP). We introduce
a variant of this type of model which maintains a reasonable model
size while addressing practically relevant problem instances. In this
way, a hybrid classical-quantum integer solver can be applied to solve
them, and a comparison with a classical solver is also possible.

In particular, we address the train rescheduling problem in complex railway networks with mixed infrastructure including single, double, and multiple-track railway lines with given planned train paths. Our consideration includes shunting movement of rolling stock between depots and stations followed by rolling stock connections. We do not include train cancellations in the model.
We apply a new hybrid quantum-classical rescheduling algorithm combining classical heuristics with Quantum Annealing (QA). This work extends on  
a particular linear modeling strategy, partly explored on a toy model by
\cite{qubohobo}.
Our (ILP) model, tailored for the particular practical railway situation, is solved with proprietary D-Wave solvers as well as with CPLEX for comparison.

We recognize several important scientific gaps which the present work aims to contribute to. First, existing optimization models do not scale well: they can become huge and inefficient for larger real-life instances.
Exploiting the characteristics of a given railway scenario, one can possibly construct simplified models whose size remains small enough to be addressed with a suitable exact or heuristic solver, yet they capture the relevant optimization objectives and constraints.
For the particular railway scenario and solver studied in the present paper, no such model is known.
Second, the QA models for the studied optimization problem that have been introduced so far had been designed for pure QA implementations and have been demonstrated only on simple network setups due to the size limitations of the currently available Noisy Intermediate Scale Quantum (NISQ) devices.
Third, to our knowledge, no hybrid QA-based models have been used for real-life railway optimization so far, nor in other schedule-based modes, like public transport or air traffic planning/rescheduling.  
In this paper, we demonstrate the quantum readiness of medium-scale
railway rescheduling models: we successfully apply quantum methods in
the rescheduling problem of an urban railway network.

The paper is organized as follows. In Section~\ref{sec:quantum} we briefly review the state of the art of quantum annealing and its applications.  In Section~\ref{sec::scenario} we describe the problem under investigation, in Section~\ref{sec::methods_models} we present our model, in Section~\ref{sec::hybrid} we discuss the hybrid solver we use, in Section~\ref{sec::computation} we present computational results. In Section~\ref{sec:conclusions} conclusions are drawn.

\section{Quantum annealing}
\label{sec:quantum}

Quantum computing devices became available for practical computational
purposes in the last few years. Quantum annealers are such devices:
they implement two-level quantum systems (quantum bits) with tunable
pairwise interactions: an Ising system, well-known in
physics~\citep{Ising_1925, Bian2010TheIM}.  The expression of the
energy of such a system is one of the many equivalent forms of quadratic
unconstrained binary optimization problems (QUBOs): the extremum of a
quadratic function of binary variables. This is an NP-hard problem in general.

The class of QUBO problems is well studied and has a number of
applications; we refer the reader to a recent monograph edited
by~\cite{punnenQUBObook}. 
Linear and quadratic constrained binary problems can also be transformed to QUBOs: linear and quadratic constraints can be taken into account with
penalties~\citep{Gusmeroli_2022}. Using appropriate binary encoding of
the variables, integer programming problems can be
treated in this way as well. Hence, an efficient QUBO solver may be useful in many
problems, including the integer linear programs arising in the present contribution.

By expressing a QUBO as the energy of an Ising system, one gets a
configuration of pairwise couplings of spins and local fields applied
to individual spins, which, in turn, can be realized in tunable quantum
hardware. According to the adiabatic theorem of quantum
mechanics~\citep{avron.elgart.99}, tuning the system's configuration
from a suitably chosen initial state to the one corresponding to the
actual objective slowly enough, the optimal
configuration can be read out under certain conditions. Unfortunately, not all QUBOs meet the
required conditions, so even in an ideal setting the quantum annealer
is not a universal solver of hard problems, nevertheless it can be efficient in
a variety of problems. Quantum annealing can be also viewed as a process
analogous to simulated annealing, in which the noise required to avoid
staying in local optima has quantum (as opposed to thermal) origins
~\citep{das_colloquium_2008}.

In real physical hardware, like the D-Wave machine, thermal noise and
imperfections cannot be avoided. Hence, the physical devices realize a
combination of thermal and quantum annealing, which renders them a
probabilistic heuristic solver for QUBOs. They output a number of configurations
typically close to, and possibly including the optimal one.

The state-of-the-art physical quantum annealers are smaller systems of
a few hundred quantum bits, where not all pairs of these are coupled. The fixed
topology of qubit couplings means that the problem's graph (defined by
the nonzero couplings) has to be embedded as an induced subgraph of the graph describing
the topology of the system. This embedding is a computationally hard problem
itself~\citep{zbinden2020embedding}. It often requires to couple
multiple physical qubits to represent a logical bit of the problem.
As quantum annealers are not algorithms running on digital computers
but analog devices implemented physically, the coefficients of the
problem can only be set with limited accuracy.

To overcome the problem of limited size and accuracy, quantum
annealers of the present state of the art are often used in hybrid
(quantum-classical) solvers, orchestrating classical algorithms and
using QA as a subroutine to address hard
problem instances more efficiently. Solvers available in D-Wave's 'Leap
Hybrid Solver Service (HSS)'~\citep{DwaveHSS}, including the one used
in the present work, belong to this family. Meanwhile, quantum
technology keeps on developing, systems of bigger size and better
topology are regularly announced, they are more and more affordable,
and there is a growing community around them.

Currently, the applicability of QA technologies is being explored in various fields. In the railway context, the first applications of QA can be found in train dispatching/rescheduling \citep{qubohobo} and rolling stock planning~\citep{2109.07212v1}. In particular, the first proof-of-concept
demonstration of a (pure) quantum computing approach to railway rescheduling was presented by some of us~\citep{domino2023quantum}. A follow-up paper by \cite{qubohobo} laid down the principles of a more general modeling approach to railway rescheduling in light of QA, introducing a suitable QUBO / HOBO (Higher-Order Binary Optimization) encoding of these problems. The present work continues this line of research, which is gaining increasing attention, see e.g. the recent work of~\cite{xu2023high}.

\section{Problem description}\label{sec::scenario}

\begin{figure}[ht]
\centering
\includegraphics[width = 0.6 \textwidth]{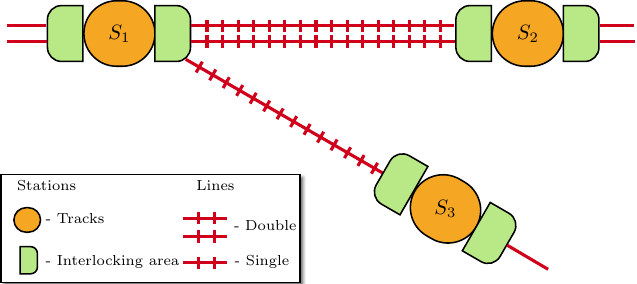} 
\caption{Edges and nodes on the example railway network. Interlocking areas of stations in green.}
\label{fig:exemplary_network}
\end{figure}

We consider a railway rescheduling problem that includes train reordering, retiming, and shunting movements with rolling stock circulation at stations.
The considered traffic takes place in an urban railway network with mixed track systems: segments of single, double, to quadruple tracks.

We model a railway network with edges and nodes of a graph. 
Figure \ref{fig:exemplary_network} depicts an example of a network layout composed of nodes (station or junction) 
and edges (single, double, multiple-track lines). 
Each edge is composed of one or more \emph{tracks}. Each track consists of 
block sections defined by pairs of signals \citep{LUSBY2013713}. Each block section can be occupied by at most one train at a time.
A subsequent train is allowed to enter
the block only after a minimal \emph{headway}: minimal time span
between the trains. To place the model into a real-life railway context, we adopt a 2-block signaling system, meaning that two free blocks are required between the consecutive trains.  The 2-block signaling system is common across the Polish railway network, and typical for conventional primary railway lines. The signaling system affects the model through the way of calculating parameter values (headway times), hence, the model can be applied to various safety systems. We assume \emph{green way policy}, that is, trains have the free way to move at the maximal allowed speed between stations. This results in constant running times for each block.

Each node is composed of station tracks (blocks) and interlocking areas. The conflicts described by our model are the following.
Each station track can be occupied by at most one train at a time.
Routing dependencies between pairs of trains competing for the same resource in interlocking areas are considered to guarantee that only one train from the pair can occupy the area at a time.

A train's \emph{route} is the sequence of blocks
the train passes during its journey.
A train \emph{path} is a sequence of arrival and departure times of a particular train assigned to a train route.
We assume a one-minute resolution for all time parameters such as timetable time, running and dwell time minimal headway time, or passing time, resulting in integer variables. 
This assumption of discretized time is needed to enable the use of quantum-based solvers, and, more importantly, introduces the possibility of pruning binary variables.
The relaxation of integer constraints on time variables is not expected to improve computation time significantly, as the real difficulty is tied to the precedence of trains which is modeled by the binary precedence variables.

In the network, two trains can follow each other keeping their given order. 
They can meet and overtake (M-O) when they go in the same direction, thereby changing their order. 
A pair of trains heading in the opposite direction can meet and pass (M-P).
On \emph{single-track} lines, M-Ps and M-Os are only possible at stations. 
This has to be enforced by constraints in our model. 
To maintain a reasonable number of constraints, the set of all pairs of trains that are involved in such constraints, that is, which can potentially meet on any single-track line segment in opposite directions, can be determined in advance from the model instance data. 

Similarly, trains in the same direction preserve their order between stations and keep the minimal headway time between each other. This does not hold for trains running in the same direction on multiple-track segments on different tracks.
The set of all pairs of trains that can potentially violate the minimal headway condition can also be predetermined, thereby minimizing the number of the respective constraints.

The train traffic is scheduled based on the given timetable. The timetable contains all the train paths. 
In the train rescheduling problem, we are given an initial timetable that is conflict-free. However, conflicts may appear due to disturbances such as late departures and/or arrivals due to excessive passenger demand, malfunctioning rolling stock, etc. 
The disturbances appear in our model via initial delays: the delays of trains at the beginning of their journey through the network.
The initial delays have fixed values in each of our problem instances; the study of the statistics of delays because of random initial delays \citep{SPANNINGER2022100312} is beyond the scope of the present study.
Following \cite{CORMAN201279},  a \emph{conflict} is an inadmissible situation when at least a pair of trains
claim the same resource (e.g., block section, switch) simultaneously. Conflicts can occur either on railway lines or at the stations. 
Possible conflicts that can occur on the lines include the lack of the minimum headway between two subsequent trains heading on the same track in the same direction, or claiming a segment of a single-track line by two trains heading in opposite directions at the same time.
Conflicts at stations include claiming a station track by two trains at the same time or claiming a station switch (in the interlocking area) by two trains at the same time.

The conflicts have to be solved by modifying the original timetable, applying decisions on the train sequencing, and retiming for trains claiming the same resource. 
Such an intervention in the structure of train paths implies additional changes to maintain the feasibility of the modified timetable. 
These changes typically result in additional delays of trains, giving rise to \emph{secondary delays} depending on the dispatching decision depending on the modification of the timetable. We will minimize a function of these secondary delays.

First we determine the subset of \emph{decision} stations: these are the nodes of the network (typically: stations) where the events that are directly affected by the decisions can take place. These events are: starting, termination, entering or leaving the modeled part of the network, meet and pass, meet and overtake, and the joining and splitting of train routes.

The choice of decision stations depends on railway operation practices and actual traffic circumstances: stations where trains are allowed to meet and pass or meet and overtake, network junctions, and the boundaries of the selected part of the network have to be considered as decision stations. 
Meanwhile, there can be stations with an infrastructure supporting meet and pass or meet and overtake, which are not commonly used for this purpose under normal circumstances. These are normally not considered as decision stations. In the case of disruptions, however, some of them can be introduced as additional decision stations upon the network operator's decision.
From the model's point of view, the decision stations are those to which decision variables are tied to. In what follows, we use the term "station" for decision stations only, unless otherwise stated.

The model does not include decision variables tied to non-decision stations. For instance, the precedence of trains cannot be changed on non-decision stations.
They do have their impact on parameter values in the model though.
Headways, for instance, are calculated between decision stations, taking into account all the line blocks and station blocks of non-decision stations in between. 

On the station where a train terminates or sets off, shunting is also modeled. The goal of shunting is to move the train from the passengers' service track to the depot or vice versa. The depot is treated as a station, and consider it as a black box, without a detailed layout. 
We treat shunting movements as a service train from the depot to the starting station of the service train, or from the terminating station to the depot. Rolling stock circulation conditions are prescribed to ensure the precedence between the service train and the actual train.

\section{Methods and Model}\label{sec::methods_models}

In the following, we describe our model in detail.
Section \ref{sec::notation} defines sets, parameters, and decision variables. Section \ref{sec::lp} describes our Integer Linear Programming (ILP)
formulation.

\subsection{Sets,  parameters and decision variables}\label{sec::notation}

In order to define our decision variables, constraints, and objective
function, we first determine sets of index tuples from the given infrastructure
data, timetable data, and the rolling stock circulation plan. Also, we
introduce parameters calculated from the same input.

\subsubsection{Sets} 
\label{sec:sets}

\begin{table}[t]
  \centering
  \begin{tabular}{|l|p{0.8\linewidth}|}
    \hline 
    $\hat{\mathcal{S}}$ & stations \\
    \hline 
    $\mathcal{S}$& decision stations \\
    \hline 
    $\mathcal{J}$ & trains \\
    \hline 
    $\mathcal{S}_j$ & stops of train $j$ (ordered) \\
    \hline 
    $\mathcal{J}^{2\,(\text{close})}$ & train pairs with precedence variables \\
    \hline 
    $\mathcal{J}^{2\,(\text{single})}$ & train pairs possibly conflicting on single-track segments \\
    \hline 
    $\mathcal{J}^{2\,(\text{headway})} $ & train pairs prone to violation of minimal headway \\
    \hline 
    $\mathcal{J}^{2\,(\text{turn})}_s$ & train pairs with the same rolling stock turning at $s$  \\
    \hline 
    $\mathcal{J}^{2\,(\text{track})}_s$& train pairs possibly competing for the same track of $s$ \\
    \hline 
    $\mathcal{J}^{2\,(\text{switch},\text{out})}_s$& train pairs possibly competing for the same switch in an interlocking area of $s$ upon departure \\
    \hline 
 $\mathcal{J}^{2\,(\text{switch},\text{out},\text{in})}_{s,s'}$ & train pairs possibly competing for the same switch in an interlocking area of $s$
  upon the departure of $j$ from $s$ while $j'$ arrives to $s$ from the direction of $s'$. \\
    \hline
    $\mathcal{J}^{2\,(\text{switch},\text{in},\text{noMO})}_{s,s'}$ & train pairs possibly competing for the same switch in an interlocking area of $s$ upon arrival from $s'$ under special circumstances described in the main text.  \\
    \hline 
    $\mathcal{J}^{2\,(\text{switch},\text{in},\text{MO})}_{s,s'}$& train pairs possibly competing for the same switch in an interlocking area of $s$ upon arrival, under special circumstances described in the main text.  \\
    \hline 
    $\mathcal{C}^2_{j}$ & station pairs subsequent in the route of $j$ \\
    \hline
    $\mathcal{C}^{2\,(\text{common})}_{j,j'}$  & station pairs subsequent in the route of both $j$ and $j'$ heading in the same direction \\
    \hline
    $\mathcal{C}^{2\,(\text{common, single})}_{j,j'} $  & station pairs which are connected with a single-track line
    segment, subsequent in the route of
    both $j$ and $j'$  heading in the opposite directions\\
    \hline
\end{tabular}
\caption{Summary of sets}
\label{tab:sets}
\end{table}

Let $\hat{\mathcal{S}}$ denote the set of all stations, and $\mathcal{S}\subset \hat{\mathcal{S}}$ the decision stations which have been selected in advance, as already described.
The main objects of our model for railway rescheduling are the
trains $j \in \mathcal{J}$ and the stations $s \in \mathcal{S}_j$ in
their path. The sets $\mathcal{S}_j$ list decision stations only and they are ordered. In addition to $\mathcal{J}$ and $\mathcal{S}_j$, the relevant sets are the following.
\begin{itemize}
\item The set $\mathcal{J}^{2\,(\text{close})}\subset \mathcal{J}\times
  \mathcal{J}$ is the set of train pairs for which a precedence variable can potentially be defined in the model. It could be considered to be the whole $\mathcal{J}\times
  \mathcal{J}$, but this would result in a significant number of redundant variables. At the description of the parameter $d_{\max}$ we will introduce a way how to initially reduce the number of train pairs potentially involved in any constraint; essentially this will be the set of trains that potentially ``interact''. This set will be divided into subsets related to specific ``interactions'' in what follows.

\item The set $\mathcal{J}^{2\,(\text{single})}\subset \mathcal{J}^{2\,(\text{close})}$ contains all train pairs that can potentially meet and pass on a single-track line segment. 
This set can be obtained by considering all pairs of trains that are heading in the opposite direction on any single-track line, assuming all possible delays. 
If their train paths possibly intersect on any single-track segment, they are in $\mathcal{J}^{2\,(\text{single})}$.

\item The set $\mathcal{J}^{2\,(\text{headway})}\subset \mathcal{J}^{2\,(\text{close})}$ contains all train pairs that can violate the minimal headway condition, that is, those which potentially follow each other heading in the same direction, assuming any of the possible delays that can occur in the model instance. 

\item The sets 
  $
  \mathcal{J}^{2\,(\text{turn})}_s\subset \mathcal{J}\times
  \mathcal{J},\ 
  \left(\forall s\in \mathcal{S}\right) $ are the sets of all pairs of trains so that the first
  train of the pair terminates at station $s$ and its rolling stock
  continues as the second train of the pair. These sets are deduced
  from the rolling stock circulation plan and the timetable. 
  
\item  The sets $\mathcal{J}^{2\,(\text{track})}_s\subset \mathcal{J}^{2\,(\text{close})},\ 
  \left(\forall s\in \mathcal{S}\right)$ are the sets of train pairs that can possibly occupy the same track on the station $s$, thereby competing for the same track.
  
\item   The sets $
  \mathcal{J}^{2\,(\text{switch},\text{out})}_s\subset \mathcal{J}^{2\,(\text{close})},\ 
  \left(\forall s\in \mathcal{S}\right)$ are the sets of
  train pairs that can possibly compete for the same switch in an interlocking area of $s$ upon their
  departure from station $s$.
  \item The sets $
  \mathcal{J}^{2\,(\text{switch},\text{out},\text{in})}_{s,s'}\subset \mathcal{J}^{2\,(\text{close})},\ \left(\forall (s,s')\in \mathcal{S} \times \mathcal{S}\right)$ are the
  sets of train pairs that can possibly compete for the same switch in an interlocking area of $s$
  in order to have $j$ to depart from $s$ while $j'$ arrive at $s$ from
  the direction of $s'$ (given trains' paths fixed, $s'$ is uniquely defined by $j'$ and $s$). For these train pairs, the respective constraints will ensure that they do not pass the same switch in an interlocking area at the same time upon departure. 
   
\item  The sets $
  \mathcal{J}^{2\,(\text{switch},\text{in},\text{noMO})}_{s,s'}\subset \mathcal{J}^{2\,(\text{close})},\ 
  \left(\forall (s, s') \in \mathcal{S} \times \mathcal{S}\right)$ are the
  sets of train pairs that can possibly compete for the same switch in an interlocking area of $s$ upon their arrival at $s$ from the direction of $s'$, and there is no M-O
  possibility for them between $s$ and $s'$.
\item The sets $
  \mathcal{J}^{2\,(\text{switch},\text{in},\text{MO})}_{s,s'}\subset \mathcal{J}^{2\,(\text{close})},\ \left(\forall (s, s') \in \mathcal{S} \times \mathcal{S}\right)$ are the sets
  of train pairs that can possibly compete for the same switch in an interlocking area of $s$
  upon their arrival at $s$ so that either both of them come from the direction
  of $s'$ but there is a M-O possibility for them between $s$ and
  $s'$, or one of them is approaching $s$ from a direction 
  other than $s'$. 
  The constraints related to these sets and those related to $\mathcal{J}^{2\,(\text{switch},\text{in},\text{noMO})}_{s,s'}$ together will ensure that pairs of inbound trains do not use the same switch in an interlocking area at the same time upon arrival, which is also enforced by the interlocking system. The possibility for M-O between the current and previous decision stations (e.g. using multiple tracks) is to be modeled specifically, hence the distinction between these sets and the $
  \mathcal{J}^{2\,(\text{switch},\text{in},\text{noMO})}_{s,s'}$-s.
  
\item The sets $\mathcal{C}^2_{j},\ \left(\forall j\in \mathcal{J}\right)$ are
  the sets of all station pairs that are subsequent in the route of
  $j$.
  
\item  The set $\mathcal{C}^{2\,(\text{common})}_{j,j'},\ \left(\forall (j,j')\in \mathcal{J} \times \mathcal{J}\right)$ are the sets of all station
  pairs that appear as subsequent stations in the route of both $j$
  and $j'$ heading in the same direction, and they cannot meet and overtake between the subsequent stations.
  
\item   The sets $\mathcal{C}^{2\,(\text{common, single})}_{j,j'},\ \left(\forall (j,j')\in \mathcal{J} \times \mathcal{J}\right)$ are those of all
  station pairs ($s$, $s'$) that appear as subsequent stations in the route of
  both $j$ and $j'$ \emph{which are connected with a single-track line
segment}, and trains are heading in opposite directions. The order within the pairs is determined by $j$. (Note that $s'$ is uniquely defined by $j$, $j'$ and $s$ because the routes of trains are fixed.)
\end{itemize}
All these sets can be enumerated based on the input data in a
straightforward manner. A brief summary of the sets is provided in Table~\ref{tab:sets}.

\subsubsection{Parameters}

\begin{table}[t]
  \centering
  \begin{tabular}{|l|p{0.8\linewidth}|}
    \hline 
    $\tau^{(\text{pass})}(j, s \rightarrow s')$ & running time of $j$ from $s$ to $s'$\\
    \hline
    $\tau^{\text{(headway)}}(j,j', s \rightarrow s') $ & minimal headway time for $j'$ following $j$ from $s$ to $s'$\\
    \hline
    $\tau^{(\text{dwell})}(s,j)$ & minimal dwell time of $j$ at $s$\\
    \hline
    $\tau^{\text{(turn)}}(s,j,j')$& rolling stock turnaround time for $j$ to continue as $j'$ from s\\
    \hline
    $\sigma(j,s)$ & originally scheduled departure of $j$ from $s$ \\
    \hline
    $\upsilon(j,s)$ & earliest possible factual departure time of $j$ from $s$\\
    \hline
    $d_{\text{max}}$ & upper bound for secondary delays\\
    \hline
\end{tabular}
\caption{Summary of parameters}
\label{tab:parameters}
\end{table}

The parameters that appear in our model, also summarized in Table~\ref{tab:parameters}, are the following:
\begin{itemize}
\item $\tau^{(\text{pass})}(j, s \rightarrow s')$ is the running time of $j$ from $s$ to $s'$.
\item $\tau^{\text{(headway)}}(j,j', s \rightarrow s') $ is the
  minimal headway time for $j'$ following $j$ from $s$ to
  $s'$.
\item $\tau^{(\text{switch})}(j, j', s)$ the time for which train $j$ occupies an interlocking area at $s$; this time should elapse before $j'$ can start occupying the same area. Our model is simplified in this respect: we consider an upper estimate of this time valid for any switching zone of the station. The assumed values are determined from the parameters of station technology and the characteristics of the train pairs.

\item $\tau^{(\text{dwell})}(s,j)$ is the minimal dwell time of $j$
  at $s$.
\item $\tau^{\text{(turn)}}(s,j,j')$ is the minimum turnaround time for the
  rolling stock of a train $j$ terminating at $s$ to continue its
  journey as train $j'$.
\item $\sigma(j,s)$ is the scheduled departure time of $s$ from $j$, according to the original timetable.
\item $\upsilon(j,s)$ is the earliest possible departure time of $j$ from
  $s$ when a disturbance in the network happens. It is the maximum of the planned departure time $\sigma(j,s)$, and the technically feasible
  earliest departure time. 
  The latter is calculated using the imaginary situation that the train starts from the initial location at the initial time defined by the initial delays (from the problem instance), and it moves to station $s$ as fast as technically possible, without having any other trains on the network, and then departs from there after the minimal stopping time has elapsed. These parameters are calculated before the optimization for the problem instances, based on the technological data of the network (minimum passing times) and the initial delays.
\item $d_{\text{max}}$ is an upper bound assumed 
for the
secondary delay $t^{(\text{out})}(j,s) - \upsilon(j,s)$. This parameter sets an upper limit for the possible secondary delays.
Setting such a bound is not uncommon in the literature, see e.g. the work of \cite{DarianoPDPbb}. 
We set the same value of this parameter for all our testing instances. It has to be big enough so that no secondary delay exceeds it (e.g. $d_{\max} = 40$ excludes the possibility of a one-hour delay due to waiting for another delayed train). 
Meanwhile, it is desirable to set its value as small as possible; restricting the time variables to small intervals decreases the number of binary decision variables in the model. 
\end{itemize}
As mentioned in Section~\ref{sec:sets}, the parameter $d_{\max}$ makes it possible to define $\mathcal{J}^{2\,(\text{close})}$, the set of trains potentially appearing in constraints, given a problem instance.
Namely, a pair of trains $(j, j')$ is included in this set if and only if they can meet at any station, given the original timetable, the initial delays, and assuming that no train can have a secondary delay greater than $d_{\max}$. The choice of a small $d_{\max}$ parameter results in a decrease of the model size, which is achieved via defining the set $\mathcal{J}^{2\,(\text{close})}$ with smaller cardinality.
Choosing too small $d_{\max}$ can result in the infeasibility of the model. 

\subsubsection{Decision variables}

\begin{table}[t]
  \centering
  \begin{tabular}{|l|p{0.8\linewidth}|}
    \hline 
    $t^{(\text{out})}(j,s)$ & departure time of train $j$ from decision station $s$\\
    \hline
    $y^{(\text{out})}(j,j',s)$ & order of trains $j$ and $j'$ upon their departure from $s$ \\
    \hline
    $y^{(\text{in})}(j,j',s)$ & order of trains $j$ and $j'$ upon their arrival from $s$ \\
    \hline
    $z(j,j',s,s')$ & the order of trains at a resource such as e.g. a single-track segment, located between $s$ and $s'$ the trains $j$ and $j'$ are competing for. \\
    \hline
\end{tabular}
\caption{Summary of decision variables}
\label{tab:variables}
\end{table}
Our decision variables, summarized in Table~\ref{tab:variables}, are the following.
First, we use departure time variables:
\begin{equation}
  \label{eq:tvars}
  t^{(\text{out})}(j,s)\in \mathbb{N}, 
 \ \ \ \text{where} \ \ \  \upsilon(j,s) \leq t^{(\text{out})}(j,s) \leq \upsilon(j,s) + d_{\max},
\end{equation}
defining the departure time of train $j\in \mathcal{J}$ from station $s\in \mathcal S_{j}$. Such a variable is defined for all decision stations $\mathcal{S}_j$. 
The bounds follow trivially from the notion of the $\upsilon(j,s)$ and $d_\text{max}$ parameters.
In the case of trains that terminate within the modeled part of the network, the last station is not included in $\mathcal{S}_j$.  
For the sake of notational convenience, we also introduce the arrival time of the trains at stations, $t^{(\text{in})}(j,s')\in  \mathbb{N}$.
Having assumed fixed running times (green way policy), arrival times are defined by the previous departure times, (c.f. Equation~\eqref{eq::min_pass_time}). 
Hence, arrival times are not independent decision variables themselves.
This explains why we need sets and decision variables related to various combinations of inbound and outbound trains when describing conflicts in switching zones.

In addition to the time variables, we use three sets of binary precedence variables. The decision variables $y^{(\text{out})}(j,j',s) \in \{0,1\}$
determine the order of trains to leave stations: such a variable takes the value $1$ if train $j$ leaves station $s$ before train $j'$, and it is 0 otherwise.
The $(j,j',s)$ tuples for which an $y$ variable is defined
will be specified later. 
Similarly, the precedence variables $y^{(\text{in})}(j,j',s) \in \{0,1\}$
prescribe the order at the entry to stations; the value is $1$ if $j$ arrives to $s$ before $j'$. 

The last set of precedence variables describes the precedence of trains $j$ and $j'$ competing for a resource located between stations $s$ and $s'$. 
The binary variable
  $z(j,j',s, s') \in \{0,1\}$
will be $1$ if $j$ uses the given resource before $j'$. 
The quadruples $(j,j',s,s')$ for which we have such a variable
will be specified later, but let us provide an example here.
Consider two stations $s$ and $s'$ that are connected by a single-track line segment, and trains $j$ and $j'$ are heading in opposite directions, competing for the single-track segment as a resource. In this case, the variable describes which of the trains can occupy the single-track link first. (Recall that M-P is possible on decision stations only.)

\subsection{ILP formulation}\label{sec::lp}

Given the index sets, variables, and the parameters and decision variables of the model, now we formulate the optimization objective and the constraints.

\subsubsection{Objective function}

Our goal is to minimize the secondary delays that occur in the analyzed part of the railway network.
A suitable objective function is the \emph{weighted sum of secondary 
delays at the destination station}:
\begin{equation}\label{eq::objective}
f(t) = \sum_{j \in \mathcal{J}} w_j  \left(t^{(\text{out})}(j, s^*) - \upsilon(j, s^*)\right),
\end{equation}
where  $s^*$ is the last element of $\mathcal{S}_j$, and $w_j$ are weights for each train representing its priority. The weights have to be chosen in such a way that they reflect the rules and practices of the local dispatching system. As these rules are often formulated only qualitatively, the particular values are somewhat arbitrary; the model can be fine-tuned in this respect. 
(In Section~\ref{sec::considered_network} we provide more details of the weight assignments for Polish PKP PLK.) 
The reason for restricting ourselves to the secondary delays at the destination stations only is that this is the actual figure of merit chosen by the respective railway authorities; the generalization to other linear objectives such as, e.g., the weighted sum of secondary delays on all stations, is straightforward.

\subsubsection{Constraints}

The constraints of the model are the following.

\paragraph{Minimal running time} Each train needs a minimal time to get to the subsequent station:
  \begin{equation}
 t^{(\text{in})}(j, s') =  t^{(\text{out})}(j, s) + \tau^{(\text{pass})}(j, s \rightarrow s'), \quad
 \forall j \in \mathcal{J}, \ \forall(s, s') \in \mathcal{C}_j.
\label{eq::min_pass_time}
\end{equation}
Recall that this condition represents the green way policy and makes the variables $t^{(\text{in})}$ depend trivially on the $t^{(\text{out})}$-s. The reason for keeping $t^{(\text{in})}$-s is that they make the model description more transparent.
\paragraph{Headways} A minimal headway time is required between subsequent train pairs on the common part of their route as
  \begin{eqnarray}
t^{(\text{out})}(j', s) \geq   t^{(\text{out})}(j, s)  + \tau^{(\text{headway})}(j, j', s \rightarrow s') - C \cdot y^{(\text{out})}(j',j,s),
\nonumber \\
\forall (j,j') \in \mathcal{J}^{2\,(\text{headway})},\ \forall (s,s') \in  \mathcal{C}^{2\,(\text{common})}_{j,j'},  
\label{eq::spacing_at_line}
  \end{eqnarray}
  where $C$ is a constant big enough to make the constraint satisfied
  whenever the binary variable $y^{(\text{out})}(j',j,s)$ takes the
  value of $1$: in this case, $j'$ goes first, so the constraint should be trivially satisfied. Otherwise speaking, the constraint should only be active for $y^{(\text{out})}(j',j,s)=0$. (Constraints of this structure are often termed as big-M constraints in this context.) 
  In order to find the smallest value of $C$ for this to hold, we can exploit the bounds on the time variables in Equation~\eqref{eq:tvars}): we have
  \begin{equation}
  t^{(\text{out})}(j', s) \geq \upsilon(j', s)
  \end{equation}
  and 
    \begin{equation}
    t^{(\text{out})}(j, s) \leq \upsilon(j, s)+d_{\text{max}},
  \end{equation}
  hence, setting
  \begin{equation}
  C = -\upsilon(j', s) + \upsilon(j, s) + d_{\max} + \tau^{(\text{headway})}(j, j', s \rightarrow s')
  \label{eq::spacing_at_line_C},
  \end{equation}
  if $y^{(\text{out})}(j',j,s)=1$ then  Equation~\eqref{eq::spacing_at_line} will take the form  
  \begin{equation}
      t^{(\text{out})}(j',s) -\upsilon(j', s) \geq t^{(\text{out})}(j,s)-d_{\text{max}}-\upsilon(j,s),
  \end{equation}
   which always holds as desired. Therefore, $C$ will be set according to Equation~\eqref{eq::spacing_at_line_C}.
   
\paragraph{Single-track occupancy} Trains moving in opposite directions cannot meet on the same single-track line segment:
  \begin{eqnarray}
t^{(\text{out})}(j', s') \geq  t^{\text{(in)}}(j, s') - C \cdot z(j',j,s',s), 
 \nonumber \\
 \forall (j,j') \in \mathcal{J}^{2\,(\text{single})}, \ \forall (s,s') \in \mathcal{C}^{2\,(\text{common, single})}_{j,j'}, 
\label{eq::single_line}
  \end{eqnarray}
  where $C$ is a big enough constant chosen similarly to that in Equation~\eqref{eq::spacing_at_line}.
  
\paragraph{Minimal dwell time} Each train has to occupy the station node for a prescribed
time duration at each station:
  \begin{equation}
 t^{(\text{out})}(j,s) \geq t^{(\text{in})}(j,s) + \tau^{(\text{dwell})}(j, s),\ \forall j \in \mathcal{J},\  \forall s \in \mathcal{S}_j. \label{eq::enter_platforms}
  \end{equation}
\paragraph{Station track occupancy} Station tracks can be occupied by at most one train at a time:
  \begin{equation}
   t^{(\text{in})}(j', s) \geq t^{(\text{out})}(j, s)  - C \cdot y^{(\text{out})}(j', j, s),\ 
   {\forall s \in \mathcal{S},\  \forall (j, j') \in \mathcal{J}^{2\,(\text{track})}_s},
\label{eq::platform2} \\
  \end{equation}
  where $C$ is chosen similarly to that in
  Equation~\eqref{eq::spacing_at_line} again.  Note that this requirement
  may not be needed for depot tracks; the exceptions can be handled
  by the proper definition of $\mathcal{J}^{2\,(\text{track})}_s$.
\paragraph{Interlocking area  occupancy} These ensure that trains that use the same switch cannot meet in interlocking  areas:
  \begin{eqnarray}
\label{eq:switch1}
t^{(\text{out})}(j', s)  \geq t^{(\text{out})}(j, s) + \tau^{(\text{switch})}(j,j', s) - C \cdot y^{(\text{out})}(j',j, s), \nonumber\\
\forall s \in \mathcal{S},\ \forall (j,j') \in \mathcal{J}^{2\,(\text{switch},\text{out})}_s;
\\
\label{eq:switch2}
t^{(\text{in})}(j', s) \geq t^{(\text{out})}(j, s) + \tau^{(\text{switch})}(j,j', s) - C \cdot z(j',j, s', s), \nonumber \\
\forall (s,s') \in \mathcal{S} \times \mathcal{S},\  \forall (j,j') \in \mathcal{J}^{2\,(\text{switch},\text{out},\text{in})}_{s,s'};
\\
\label{eq:switch3}
  t^{(\text{in})}(j', s) \geq  t^{(\text{in})}(j, s) + \tau^{(\text{switch})}(j,j', s) - C \cdot y^{(\text{out})}(j',j, s'), 
  \nonumber \\
  \forall (s,s') \in \mathcal{S} \times \mathcal{S},\  \forall (j,j') \in \mathcal{J}^{2\,(\text{switch},\text{in},\text{noMO})}_{s,s'};
  \\
  \label{eq:switch4} 
 t^{(\text{in})}(j', s) \geq   t^{(\text{in})}(j, s) + \tau^{(\text{switch})}(j,j', s) - C \cdot y^{(\text{in})}(j',j, s), \nonumber \\
 \forall (s, s') \in \mathcal{S} \times \mathcal{S},\  \forall (j,j') \in \mathcal{J}^{2\,(\text{switch},\text{in},\text{MO})}_{s,s'},
  \end{eqnarray}
  with a choice of $C$ similar again to that in
  Equation~\eqref{eq::spacing_at_line}. These constraints are also necessary for to make the realization of the modified timetable possible using the interlocking systems.

\paragraph{Rolling stock circulation constraints} These are introduced to bind a train with a shunting movement, termed also as a service train. If the train set of train $j$ which terminates at
  $s$ is supposed to continue its trip as (service train) $j'$ or vice versa, a precedence of these trains including a minimum turnaround time has to be ensured:
  \begin{equation}
   t^{(\text{out})}(j', s) \geq t^{(\text{in})}(j, s) + \tau^{\text{(turn)}}(j,j', s),\ 
   \forall s \in \mathcal{S} \forall (j,j') \in \mathcal{J}^{2\,(\text{turn})}_s.
\label{eq::rolling_stock}
  \end{equation}
\paragraph{Order of trains} We have additional conditions on the $y$-variables, concerning the case when M-O is not possible on the line or station. In particular, assume that the trains $j$ and $j'$ are following each other from $s$ to $s'$, heading in the same direction. If they use the same track on $s'$, they will have to leave $s'$ in the same order as they left $s$. Formally:
\begin{equation}
\begin{split}
y^{(\text{out})}(j,j',s) = y^{(\text{out})}(j,j',s'),\\ 
\forall (j_0,j_0') \in \mathcal{J}^{2\,(\text{headway})},\ \forall (s,s') \in  \mathcal{C}^{2\,(\text{common})}_{j_0,j_0'}  \\ \forall  (j, j') \in \mathcal{J}^{2\,(\text{headway})} \cap \mathcal{J}^{2\,(\text{track})}_{s'}  
\end{split}
\label{eq::order1}
\end{equation}
Similarly, if trains $j$ and $j'$, enter $s'$ from any location, they are passing the switching zone of station $s'$, then they are passing the station $s'$ in the same direction using the same track, they have to keep their order upon departure. Formally:
\begin{equation}
\begin{split} 
 y^{(\text{in})}(j,j',s') = y^{(\text{out})}(j,j',s'),\\
\forall (s, s') \in \mathcal{S}\times S,\  
\forall (j,j') \in \mathcal{J}^{2\,(\text{switch},\text{\text{in}},\text{MO})}_{s',s} \cap \mathcal{J}^{2\,(\text{track})}_{s'}.
\end{split}
\label{eq::order2}
\end{equation}
The objective function in Equation~\eqref{eq::objective}, together with the
constraints in Equation~\eqref{eq::min_pass_time}-\eqref{eq::order2} define
our mathematical programming model. 

Our model is formulated as an integer linear program. Moreover, as the time variables are constrained a finite range using the parameter $d_{\max}$, all variables are of a finite range.
We intend to limit the number of variables as much as possible. The binary variables have the obvious symmetry property
\begin{equation}\label{eq::y_eq}
\begin{split}
y^{(\text{out})}(j',j,s) = 1 - y^{(\text{out})}(j,j',s) \\
  z(j',j,s', s) = 1 - z(j,j',s, s').
\end{split}
\end{equation}
These are added to the model as constraints, in a way that only half of the variables are considered as independent, and the rest are removed.

\subsubsection{Scaling analysis}\label{sec::scaling}
Let us discuss the number of variables and constraints in our model. To assess the problem's size we will use here a worst-case approximation, namely, we assume temporarily that all trains visit all stations and any pair of trains can meet each other; these will be assumed throughout the entire consideration on scaling. Under such assumptions the number of time variables $\# t$ can be upper bounded by the product $\#\mathcal{J}\#\mathcal{S}$ of the number of trains and the number of stations:  
\begin{equation}
\# (t)  \leq \#\mathcal{J}\#\mathcal{S} .
\label{eq::no_tvars}
\end{equation}
 Each train meets at most $(\# \mathcal{J} - 1) \# \mathcal{S}$ trains going in same direction, which results in $\# \mathcal{J} (\# 
 \mathcal{J} - 1)   \# \mathcal{S} $ $y^{(\text{out})}$ variables in the model. 
 However, because of the symmetry property in Equation~\eqref{eq::y_eq}, only half of these variables are independent, thus we have at most $  \frac{\# \mathcal{J}(\#\mathcal{J} - 1) \cdot \# \mathcal{S}}{2}$ of these variables. 
 As for the trains going in opposite directions, each train meets $\# \mathcal{J} - 1$ trains at each station.
 (Although we have a quadruple of indices $(j,j', s, s')$ in the definition of $z$ variable, as routes of trains are fixed within each model instance, the elements of this quadruple are not independent.
 In particular, the index $s'$ is uniquely defined by the other indices). Hence, the number of $z$ variables is at most $\frac{\# \mathcal{J} (\# 
 \mathcal{J} - 1) \cdot  \# \mathcal{S}}{2} $ (having taken into account the  symmetry property in
Equation~\eqref{eq::y_eq} again).
Variables $y^{(\text{in})}$ occur mainly in Equations~\eqref{eq:switch1}-\eqref{eq:switch4}, and are relevant relevant for some stations only.
Nevertheless, in the worst case there can be one such variable assigned for each pair of trains and each station, yielding $  \frac{\# \mathcal{J}(\#\mathcal{J} - 1) \cdot \# \mathcal{S} }{2}$  $y^{(\text{in})}$ variables after the reduction by symmetry properties.
In summary, the upper bound for the number of precedence variables reads
\begin{equation}
\begin{split}
\# y^{(\text{out})}  &\leq  \frac{\# \mathcal{J} (\# 
 \mathcal{J} - 1) \cdot  \# \mathcal{S}}{2}
\\  \# y^{(\text{in})} &\leq \frac{\# \mathcal{J} (\# 
 \mathcal{J} - 1) \cdot  \# \mathcal{S}}{2}
\\ \# z &\leq  \frac{\# \mathcal{J} (\# 
 \mathcal{J} - 1) \cdot  \# \mathcal{S}}{2}
\\ \# y + \# z &= \# y^{(\text{out})} + \# y^{(\text{in})} + \# z \leq  \frac{3 }{2} \#\mathcal{J} (\#\mathcal{J} - 1) \#\mathcal{S} 
\label{eq::no_y_z}
\end{split}
\end{equation}
Note that these bounds are rather pessimistic: 
setting the parameter $d_{\max}$ as an upper bound on the secondary delays implies a reduction in the number of trains that can potentially conflict. 
However, the actual number of them can depend on the train density and structure of the timetable in various ways, hence, it is not possible to provide a tighter bound that would be valid in general. 
 Finally, let us remark that as the $z$-variables are mainly tied to single-track traffic, the model is more complex for networks dominated by single-track segments.
Based on Equations~\eqref{eq::no_tvars} and~\eqref{eq::no_y_z},  the scaling of the upper bound on the total number of decision variables can be estimated by
\begin{equation}
\# vars = \# t + \# y + \# z \approx O (\#\mathcal{J}^2 \#\mathcal{S}),
\label{eq::no_vars}
\end{equation}
which is quadratic in the number of trains and linear in number of stations. 

Now let us estimate the number of constraints. Considering the headway constraints in Equation~\eqref{eq::spacing_at_line}, there are at most two such constraints for each independent $y^{(\text{out})}$ variable. (Because of the symmetry property in Equation~\eqref{eq::y_eq}, only half of the $y^{(\text{out})}$ variables are independent). 

Along the same arguments, in case of the constraints in Equation~\eqref{eq::platform2}, i.e. station track occupancies, there are at most $2$ of these for each independent $y^{(\text{out})}$ variable. Analogously, for the single-line track occupancy constraints in Equation~\eqref{eq::single_line} there are at most $2$ constraints per each independent $z$ variable. 
Furthermore, in the worst case there are: $2$ constraints for each independent $y^{(\text{out})}$ variable from Equation~\eqref{eq:switch1}, $2$ constraints for each independent $y^{(\text{out})}$ variable from Equation~\eqref{eq:switch3}, $2$ constraints for each independent $z$ variable from Equation~\eqref{eq:switch2}, and $2$ constraints for each independent $y^{(\text{in})}$ variable from Equation~\eqref{eq:switch4}. Further, we expect one minimal dwell time constraint per train and station, resulting in $\#\mathcal{J}\#\mathcal{S}$ constraints in the worst case.
Concerning train order constraints in Equations~\eqref{eq::order1} and~\eqref{eq::order2}, there can be one constraint per train and station, which yields additional $ \#\mathcal{J}\#\mathcal{S}$ constraints in the worst case. 
As we do not consider splitting and joining trains on the route, at most one rolling stock circulation constraint arises per train, yielding at most $\# \mathcal{J} $ such constraints.
In summary, the total number of constraints obeys the following (worst-case) upper bound:
\begin{equation}
\# constr. \leq 8\#y^{(\text{out})} + 2\#y^{(\text{in})} + 4\#z + 2\#\mathcal{J}\#\mathcal{S} + \#\mathcal{J}.
\label{eq::no_contraints_general}
\end{equation}
Substituting the estimated number of variables in Equation~\eqref{eq::no_y_z}  we have:
\begin{equation}
\# constr. \leq 7 \# \mathcal{J}^2 \# \mathcal{S} - 5  \#\mathcal{J} \#\mathcal{S}  + \# \mathcal{J}.
\label{eq::no_contraints_limit}
\end{equation}
Let us stress that the provided generic pessimistic upper bounds only serve
the purpose of getting an overall picture of the scaling; the
particular number of variables and constraints is lower and will be
specified for our problem instances.

\section{Hybrid quantum-based approach}\label{sec::hybrid}

To overcome the limitations of current hardware quantum annealers
described in Section~\ref{sec:quantum}, we apply a hybrid
(quantum-classical) solver. In particular, we use the 'Leap Hybrid
Solver Service (HSS)'~\citep{DwaveHSS}; a cloud-based proprietary
solver that is developed by the market leader of QA hardware and is
available as a service. It implements a hybrid approach that combines
classical computational power with quantum processing.
In particular, we have used the Constrained Quadratic Model
(CQM)~\citep{dwaveCQM} Solver. The CQM hybrid solver works as follows.
\vspace{-0.4cm}

\paragraph{Input} A constrained linear or quadratic
model, and the value of the solver parameter {\tt t\_min}, which prescribes the minimum required run time (in seconds) the solver is allowed to work on any problem.
\paragraph{Preprocessing} Preprocessing includes subproblem identification and decomposition to smaller instances~\citep{tran2016hybrid}. 
Furthermore, the solver's internal preprocessing mechanism takes constraints into account
automatically, implementing the suitable penalty method internally.
\paragraph{Processing} The hybrid
solver implements a workflow that combines a portfolio of various
classical heuristics (including tabu search, simulated annealing,
etc.) running on powerful CPUs (Central Processing Units) and GPUs (Graphics Processing Units). In the course of the solution
process the hardware quantum annealer is also invoked, to
solve smaller QUBO
problems that can potentially boost the classical heuristics. Integer
problems are transformed to binary using state-of-the-art methods like
the one described by \cite{glover2019quantum,karimi2019practical}, generally applicable
for integer problems. When using the quantum hardware, the solver
performs the required embedding, runs the QUBO subproblem on the
physical QA device several times, and returns a sample
of potential solutions. After these readouts, the sample is
incorporated into the solution workflow. In this way, even though the
physical annealer supports problems with limited size (e.g. the
Pegasus machine has $5500$ physical qubits~\citep{dattani2019pegasus},
and several of these may be needed to represent a binary variable
because of the embedding), the approximate solutions of small
subproblems contribute to the solution process. The final solution is assembled by the solver's module.
\paragraph{Output} The solution or solutions of quadratic constrained problem or the integer problem.  The workflow results
in a solution or solutions that are not guaranteed to be optimal, hence, the hybrid
solver as a whole is in nature a heuristic solver. Furthermore, the solver is probabilistic, hence at each run the final output can be different. 
The CQM (integer) solver returns a series of solutions, which can be an additional asset.
For the probabilistic nature of the solver, the returned set of solutions is often termed as a \emph{sample}, and the process of obtaining the sample is termed as \emph{sampling}.
Note that CQM, like most heuristic solvers, does not warrant feasibility by default, hence, a direct check is performed after obtaining the possible solutions.
Nevertheless, for the sake of presentation, our approach is to analyze only the best CQM solution given each run of the hybrid algorithm (realization).

While D-Wave's proprietary solvers act as black boxes, i.e. they hide the exact details of the described process from the user, the output is supplemented with
timing parameters, including \emph{run time}: the total elapsed time
including system overhead, and \emph{Quantum Processing Unit (QPU) access time} which is the
time spent accessing the actual quantum hardware. These parameters
enable a comparison with other solvers.

As mentioned before, in addition to the actual problem to be solved, the CQM solver optionally inputs another
parameter, {\tt t\_min}. 
This is a time limit for heuristics running in parallel\citep{dwaveCQM}: unless a thread does not terminate by itself after {\tt t\_min}, it is allowed to run at least till {\tt t\_min} before it is stopped at the current best solution.  
We have sampled the CQM problems with various settings of {\tt t\_min} and
uncovered the impact of this setting on the model performance
and solution quality for our
problem. 

The CQM Solver can solve problems encoded in the form of Equation~\eqref{eq::objective} -- \eqref{eq::y_eq} (c.f. Section~\ref{sec::lp}) with up to $5000$ binary or integer variables and $100,000$ constraints.
Computational results of railway rescheduling problems were obtained using classical as well as hybrid (quantum-classical) solvers.  

As a classical solver, we have opted for IBM ILOG CPLEX Interactive Optimizer (version 22.1.0.0) as one of the most prevalently used commercial solver. The CPLEX computation was performed on $16$ cores of \textit{Intel(R) Core(TM) i7-10700kF CPU 3.80GHz} with $64$GBs of memory.  As a hybrid solver, the D-Wave CQM hybrid solver (version: 1.12) was used.
In addition to solving the problems to optimality, we include CPLEX results with limited computational time, for the sake of a fair comparison with heuristics.

\section{Computational results}\label{sec::computation}

In this section, we demonstrate the performance of the proposed hybrid (quantum-classical) approach on a network of Polish railways - the central part of the Upper Silesia Metropolis.
In particular, we compare the performance of D-Wave's hybrid quantum-classical solver (CQM) with CPLEX. 
The comparison is performed 
for diverse network layouts including single-, double-, and multiple-track lines, and different levels of disturbances including various initial delays and track closures. The considered closures are situations in which one or more tracks between decision stations are closed and the traffic has to be redirected to the remaining tracks between decision stations. This can imply the need to add extra decision stations to the model to bind the part of the line with the closure; this also occurs in our examples. 

Section~\ref{sec::considered_network} describes the considered network and traffic characteristics.
Section~\ref{sec::computation_generic} presents results concerning generic synthetic examples on the core line of the network, comparing double-track-line scenarios with single-track-line scenarios and closure scenarios.  Section~\ref{sec::results_real_network} presents similar results for larger network with real-life timetable. Here the case studies concern various scenarios with different degrees of difficulty.

\subsection{Railway network}\label{sec::considered_network}

\begin{figure}[ht]
\centering
\includegraphics[width = 0.6 \textwidth]{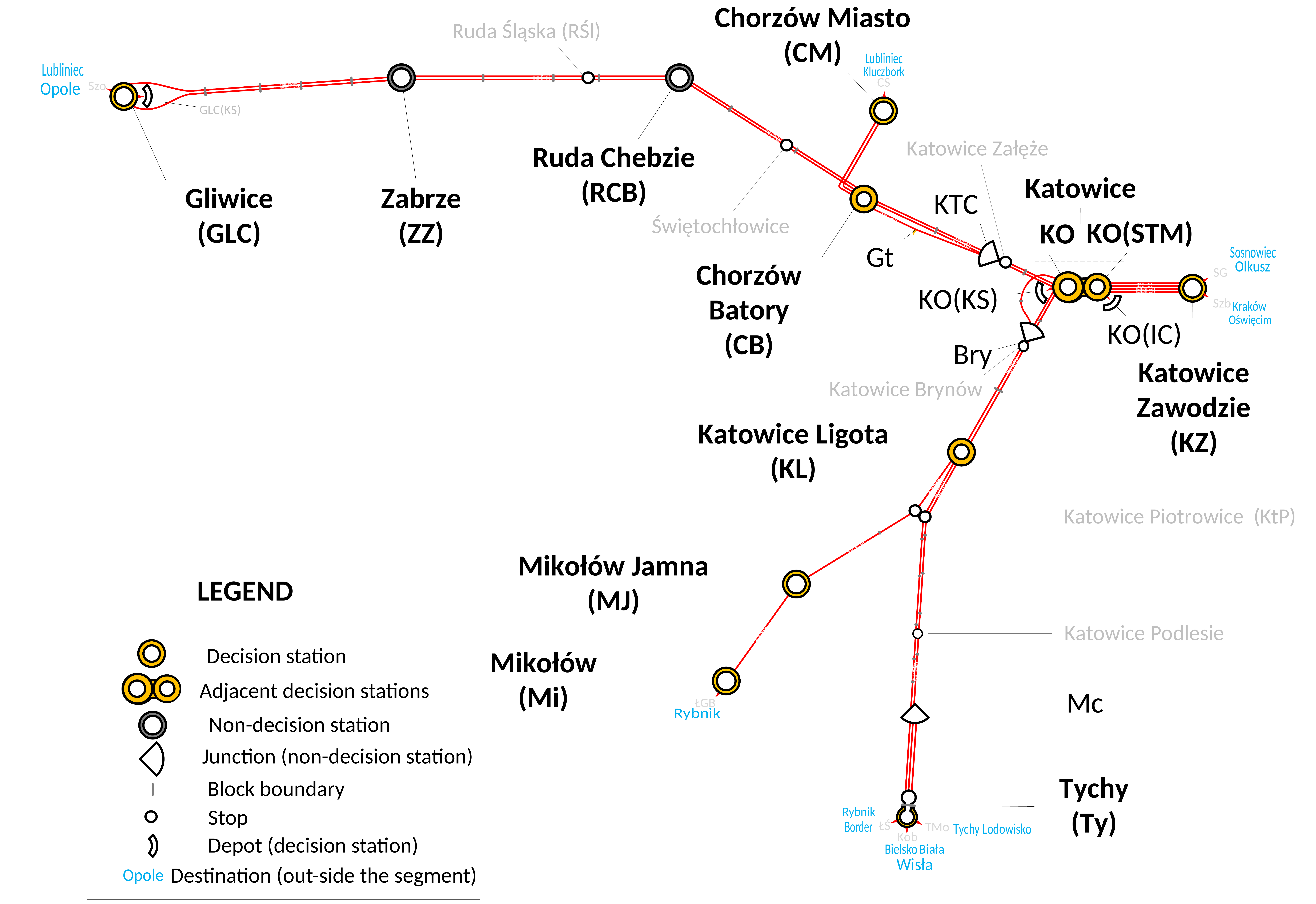}
\caption{The part of the Polish railway network which is the subject of our considerations. It is located in the central part of the Metropolis. The displayed non-decision stations such as ZZ or RCB become decision stations in certain problem instances, e.g. due to an assumed closure of one of the tracks between ZZ and RCB.}
\label{fig::network}
\end{figure}

We use the Open Railway Map\footnote{\url{https://www.openrailwaymap.org/}, visited 2022-11-25} representation of the selected part of the Polish railway network located in the central part of the Metropolis GZM (Poland), presented in Fig~\ref{fig::network}. The selected part of the network combines a heterogeneous network layout including single-, double-, triple-, and quadruple-track lines.  It comprises 25 nodes including $11$ stations and $3$ junctions, $146$ blocks, and 2 depots. 
The infrastructure is managed by Polish state infrastructure manager PKP PLK (Polskie Koleje Państwowe, Polskie Linie Kolejowa eng. Polish State Railways, Polish Railway Lines).

The decision stations include stations \{KO, KO(STM), CB, KL\}, stations bounding the analyzed part of the network \{GLC, CM, KZ, Ty, Mi\}, depots \{KO(KS), KO(IC)\} from which regional and intercity trains are being shunted to and from \{KO\}, and stations on the single-track line -- \{MJ\} to allow meet and pass (M-P) or meet and overtake (M-O).
As described before, the decision stations are the only ones that appear explicitly in our model by assigning decision variables.
Due to operational reasons resulting from track design, the decisions on train departure times (and thus on trains' order) are, in the current operational practice, \emph{de facto} made concerning the decision stations.
All other stations and junctions are treated as line blocks as long as no rerouting or retracking (e.g. in the case of closure) is considered; they will appear in the model implicitly via parameters. In such a case precedence of trains cannot be changed on these stations as there are no decision variables assigned to them.

These stations include:
\begin{itemize}
\item stations $\{ \text{ZZ}, \text{CB} \}$ in which usually there are rigid  assignments of the platform tracks to the traffic direction (i.e., track $1$ towards $\{\text{GLC} \}$; track $2$ towards $\{ \text{KO} \}$)
\item branch junctions $\{\text{KTC}, \text{Bry}, \text{Mc} \}$ in which usually there is a rigid assignment to the tracks.
\end{itemize}

As an example, let for a particular train $j$, $\hat{\mathcal{S}}_j = (\text{KZ, KO(STM), KO, KTC, CB, RCB, ZZ, GLC})$ be the ordered set of stations. Then, the dispatching decisions are implemented at stations $\mathcal{S}_j = (\text{KZ, KO(STM), KO, CB, GLC})$. The set of decision stations can be extended if necessary.

The priority weights in Equation~\eqref{eq::objective} are set to
$w_j = 1$ for stopping (local) trains 
$w_j = 1.5$ for intercity (fast) trains 
$w_j = 1.75$  for express trains, and
$w_j = 0$  for shunting (if applicable), in all the computations. These weights result from interpreting the instructions on train traffic management rules and priority levels for trains set out in the ''Instruction for train operation (Ir-1)''
\footnote{see the paragraph $51$ of Instrukcja o prowadzeniu ruchu pociagow Ir-1 (in Polish), Eng. Instruction for railway traffic \url{https://www.plk-sa.pl/files/public/user_upload/pdf/Akty_prawne_i_przepisy/Instrukcje/Wydruk/Instrukcja_o_prowadzeniu_ruchu_pociagow_Ir-1_-_wchodzi_w_zycie_od_14_sty....pdf} accessed 18 November 2024}
and the real-life dispatching practices on the PKP PLK network. The default value for regular traffic is $w_j = 1$, while there is no priority for shunting. Observe the higher priorities of InterCity trains and the even slightly higher priorities of express trains operated by Express Intercity Premium bullet trains which are briefly referred to as ``express trains''. These intercity trains and express trains have strong priority in the case of small-scale disturbances, but this priority can be neglected in the case of larger disturbances.
As for the solver parameter {\tt t\_min}, after initial experiments, we have decided to prefer {\tt t\_min}$=5$s as it provides a good balance between computation time and expected solution quality. A sensitivity analysis regarding this parameter will be performed in Section~\ref{sec::results_real_network}.

\subsection{Synthetic experiments on the selected railway line}\label{sec::computation_generic}

As the first set of experiments we address, synthetic instances on a part of the network - railway line KO-GLC, see Fig~\ref{fig::network}.
The goal in these scenarios is to compare the CPLEX and CQM solvers for: the double-track line ($line1$), the double-track line with closures ($line2$), and the single-track line  ($line3$). Shunting movements and rolling stock circulation are not considered in this set of experiments.
In particular, the addressed configurations are the following:
\begin{enumerate}
\item $line1$, the double-track line with dense traffic (As illustrated in Fig~\ref{fig::network}, we have $3$ decision stations: KO, CB and GLC). We use 3 periods of an hourly repeating timetable with $10$ trains each hour and in each direction. Here, and also in the following lines we did not consider one of the trains for practical reasons, thus we have $59$ trains in a three-hour time horizon.
\item $line2$, similar to $line1$, with the additional closure of one of the tracks between ZZ and RCB, hence, these stations are added to the set of decision stations, and single-track traffic is assumed between them.
We have $39$ trains on a two-hour time horizon. 
\item $line3$, the whole line is considered a single track. As in the "$line2$" case, we add two extra decision stations, ZZ and RCB where M-P and M-O are possible. We have $21$ trains in a three-hour time horizon. 
\end{enumerate}
For $line1$ and $line3$, the original timetable would be feasible without the initial delays, while for $line2$ it is not feasible due to the closure. 
The instances are summarized in Tab~\ref{tab::instances_generic}.
\begin{table}
\centering
\begin{tabular}{|l|l|l|l|l|}
\hline
line & $\# J$ & n.o. hours & decision stations    & track layout  \\ \hline
1    & $59$   & $3$ & KO, CB, GLC   & double  \\ \hline
2    & $36$   & $2$ & KO, CB, RCB, ZZ, GLC & \begin{tabular}[c]{@{}l@{}}single between RCB and ZZ\\ then double\end{tabular} \\ \hline
3    & $21$   & $3$ & KO, CB, RCB, ZZ, GLC & single  \\ \hline
\end{tabular}
\caption{Features of synthetic instances.}\label{tab::instances_generic}
\end{table}

For each of the lines, we compute $12$ instances, each with different initial delays of train subsets. 
The initial delays were assigned manually, a-priory for each instance, choosing plausible values based on everyday practice. Instance $0$ has no initial delays. 
Thereafter, we use the parameter value $d_{\max} = 40\ \text{min}$, in all cases in this and subsequent subsection.  
We chose its particular values by experimentation.

The upper limits found in Section~\ref{sec::scaling}  are compared to the actual number of variables and constraints (mean over instances) in Table~\ref{tab::no_vars_generic}. The purpose of the table is to connect the theoretical discussion on scaling in Section~\ref{sec::scaling} with the actual problem sizes.
For example, $line2$ is the most complex in terms of constraints, which is reflected later by the computational difficulties of some of its instances. (Due to closure, $line2$ is partially double-track and partially single-track with heavy traffic.). The table serves also as the sanity check of our model.

 \begin{table}
 	\centering
\begin{tabular}{|l|l|l|ll|ll|ll|}
\hline
\multirow{2}{*}{$line$} & \multirow{2}{*}{\begin{tabular}[c]{@{}l@{}}  {$\#\mathcal{S}$}  \end{tabular}} & \multirow{2}{*}{\begin{tabular}[c]{@{}l@{}}  {$\#\mathcal{J}$}\end{tabular}} & \multicolumn{2}{l|}{ \#  int vars  }  & \multicolumn{2}{l|}{ \#  precedence vars}     & \multicolumn{2}{l|}{ \#  constraints}    \\ \cline{4-9}   &     &  & \multicolumn{1}{p{0.5cm}|}{actual mean} & \begin{tabular}[c]{@{}l@{}}Equation~\eqref{eq::no_tvars}\\ upper lim\end{tabular}  & \multicolumn{1}{p{0.5cm}|}{actual mean} & \begin{tabular}[c]{@{}l@{}}Equation~\eqref{eq::no_y_z}\\ upper lim\end{tabular}  & \multicolumn{1}{p{0.5cm}|}{acctual mean} & \begin{tabular}[c]{@{}l@{}}\begin{tabular}[c]{@{}l@{}}Equation~\eqref{eq::no_contraints_limit}\\ upper lim\end{tabular}
\end{tabular} \\ \hline
1   & 3   & 59     & \multicolumn{1}{l|}{118 \ \ \  }  & 177     & \multicolumn{1}{l|}{1538 \ \ } & 15399  & \multicolumn{1}{l|}{7889 \ \ \ } & 72275  \\ \hline
2   & 5   & 39     & \multicolumn{1}{l|}{142}     & 195     & \multicolumn{1}{l|}{1394}     & 
11115  & \multicolumn{1}{l|}{8491}     & 52299\\ \hline
3   &   5  & 21 & \multicolumn{1}{l|}{79}     & 105  & \multicolumn{1}{l|}{592}     & 3150 & \multicolumn{1}{l|}{3057}     &  14931\\ \hline
\end{tabular}
\caption{Estimated and actual problem sizes on $line1$ $line2$ and $line3$ scenarios. For each scenario, the mean over instances was computed. }\label{tab::no_vars_generic}
\end{table}

\begin{figure}[tb]
\centering
\includegraphics[width = 0.31 \textwidth]{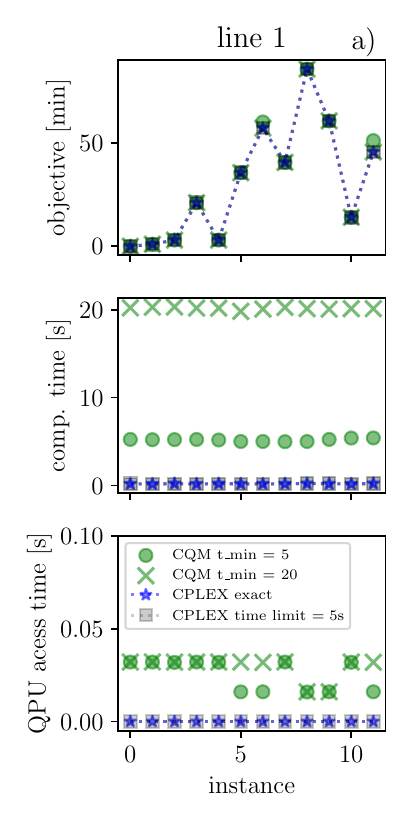}
\includegraphics[width = 0.31 \textwidth]{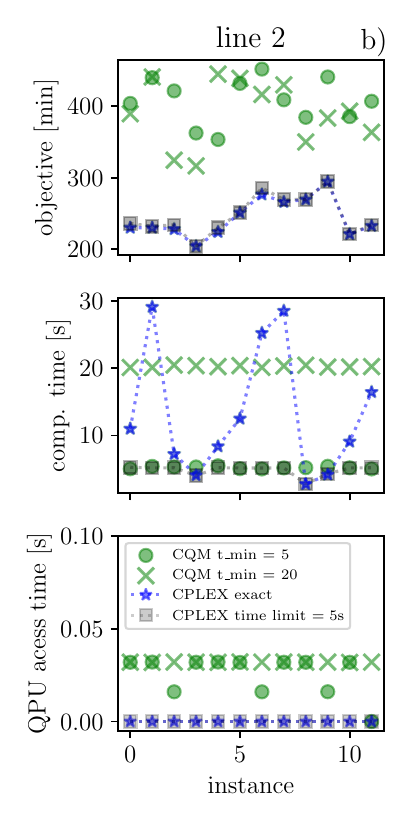}
\includegraphics[width = 0.31 \textwidth]{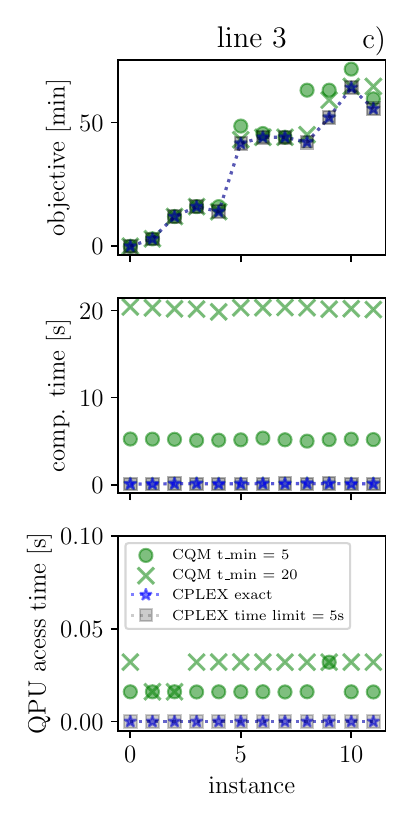}
\caption{Synthetic experiments $line1$, $line2$, $line3$. Comparison of the performance of classical solver (CPLEX) with that of the hybrid CQM. All the displayed solutions are feasible. Total computational time (middle panel) and QPU times (lower panel) were provided by the D-Wave output. Panel a) - $line1$, results of both solvers were similar in terms of the objective, but the CPLEX computation was faster. Panel b) - $line2$ namely the double-track line with closures and dense traffic, has appeared to be most challenging for both the classical solvers (in terms of computational time) and the quantum solver (in terms of objective). 
In this example, there are instances where the current CQM D-Wave hybrid solver provides feasible, though not optimal solutions of similar quality to those that CPLEX provides when its computational time is limited to be similar to that of CQM.
When CPLEX is required to solve the problem to optimality, it can be slower than CQM. 
Increasing the value of the {\tt t\_min} parameter does, in most cases, improve the quality of the solution at the cost of computational time.
Panel c) - $line3$ all the displayed solutions are feasible. The results of both solvers were similar in terms of the objective, but CPLEX computation was faster.
}\label{fig::ko-glc-case1}
\end{figure}

The computational results are presented in Figure~\ref{fig::ko-glc-case1} a) b) c) for $line1$, $line2$, and $line3$, respectively. Each figure shows the objective (top part), computation time (comp. time, middle part), and QPU access time (bottom part) for CQM hybrid solver with {\tt t\_min} equal $5$ and $20$s. CPLEX results are also displayed. These latter include the optima as well as the heuristic results with computational time limited to at most 5 seconds. 
 
For the scenario $line1$ (Figure~\ref{fig::ko-glc-case1} a) ), the CQM solver found the optimal solutions in all of the cases except for instances $6$ and $11$.
Similarly, for $line3$ (Figure~\ref{fig::ko-glc-case1} c) ), only small deviations from the optimum can be observed for a number of instances. 
All computational times for $line1$ and $line3$ were significantly below $30$seconds, although CPLEX was faster, and found the actual optimum before the possible 5 seconds limit.
Comparing CQM with {\tt t\_min}$=5s$ and {\tt t\_min}$=20s$, we have almost the same objective values, however, {\tt t\_min}$=20s$ requires almost $4$ times longer computational time. 

For $line2$ (Figure~\ref{fig::ko-glc-case1} b) ), all CQM solutions were suboptimal, although feasible. 
More remarkably, for $9$ instances at {\tt t\_min}$=5s$  and $3$ instances at {\tt t\_min}$=20s$ (out of $12$ instances) CQM  was faster than exact CPLEX, in several cases up to $5$ times faster.
CPLEX could not solve these instances to optimality when it was limited to 5 seconds; it returned close-to-optimal feasible solutions though.
This difference in time is clear while setting the minimal processing time of the CQM solver {\tt t\_min}$=5s$ (see Section~\ref{sec::hybrid}).
In principle, $line1$ (Figure~\ref{fig::ko-glc-case1} a) ) and $line3$ (Figure~\ref{fig::ko-glc-case1} c) ) are less complicated in terms of number of constraints, and here solutions of CPLEX and CQM are similar in terms of objective, but CPLEX is faster. We can conclude that the utility of the CQM solver can be expected for the more complex instances of railway scheduling problems, as the present instances for $line2$.

The detailed analysis of the implications of suboptimality in CQM solutions for $line2$ from the railway point of view shows that there are two types of consequences: the increase of the passing time between subsequent nodes, and the change in certain trains' order. Concerning the first issue, the realization of such suboptimal solutions can result in slowing down trains, which can be a motivation for further research, such as in the work of~\cite{domino2024baltimore}. 
Solutions affected by the second issue are suboptimal but still feasible and can be opted for by the human dispatcher because they have benefits for certain non-quantified reasons. 
Particular examples of real-life problem instances are presented in Appendix A.

The hybrid quantum-classical solver CQM does not outperform purely classical heuristics in most problems at the present state of the art, and railway dispatching is not yet an exception. 
Our results demonstrate, however, that it already has the potential to handle difficult railway problems of moderate size.
This implies that CQM, as a generic out-of-the-box solver, could already be a viable alternative to the currently existing solvers. Note that suboptimality can be accepted in many cases as it will be justified by the analysis of our real-world examples in Section~\ref{sec::results_real_network}.
Further improvement in the runtime and the solution quality can possibly be achieved by developing custom classical-quantum heuristics tailored for the given problem, providing also information on when the quantum solvers can be the most efficient.
Meanwhile, as there is a significant R\&D effort put into the development of quantum hardware to get beyond the NISQ era, custom algorithms, and even generic hybrid solvers can become more competitive in the future. This may render quantum approaches important or even dominant for more complex disruptions and on larger scale.

\subsection{Real-life experiments on the considered rail network}\label{sec::results_real_network}

In this subsection, we will test the CQM solver on more realistic scenarios of the railway traffic on the presented network. 
We will also focus on the track closure situation, turning a double-track line segment into a single-track one, under dense traffic, to elaborate our findings from $line2$ in Section~\ref{sec::computation_generic}.
Our computations address various use cases of train delays outside the network and within the network as well as the track closures. We constructed $9$ networks with increased levels of difficulty ranging from small delays to disruptions, i.e. it is expected more trains are involved and/or disturbances are spread more over the network. Delays were assumed a priority fixed for each network manually according to everyday practice. These networks are tabulated in Table ~\ref{tab::cases}.
Their details are as follows:
\begin{enumerate}
\item Networks $1$ - $3$ concern only disturbances due to delayed trains and no closures. 
\item Networks with number $4$ or higher also include disturbances due to closures. 
\begin{enumerate}
\item Networks $4$ and $5$ concern rerouting trains from the double-track line KTC-CB to a single-track line with higher passing times, see Figure~\ref{fig::network} (trains have no initial delays in network $3$ and some initial delays in network $4$). 
\item Network $6$ concerns multiple closures, i.e. change of multiple line KZ-KO(STM) and double-track line KO-KL to single-track lines.
(See Figure~\ref{fig::network}; the reason can be e.g. upcoming reconstruction works on this part of the network.)
\item Networks $7$ - $9$ concern closures of both $5$ and  $6$, but each with different initial delays of trains. Our intention with the last $3$ networks is to create a truly challenging rescheduling problem. 
\end{enumerate}
\item For comparison, network $0$ is the default problem with no disturbances.
\end{enumerate}
\begin{table}[!tb]
\centering
\begin{tabular}{|l|l|l|l|l|}
\hline
\makecell{network} & $\# S$& \makecell{place} & \makecell{trains\\ involved}& \makecell{description}\\ \hline
\multicolumn{5}{|c|}{\textbf{A. Disturbances from outside the network}}\\ \hline
1 & 10   & KO-Ty  & \makecell[c]{IC 14006\\ KS 94113} & \makecell[l]{Delayed IC (higher priority) \\ is in conflict with on-time KS}\\ \hline
2   & 10 & arr. Ty & \makecell[l]{KS 94766\\ KS 40518\\ IC 41004\\ KS 44862\\ IC 4120} & \makecell[l]{Delayed 5 trains \\ on the arrival to Ty, arbitrary\\  delays of several minutes}\\ \hline
\multicolumn{5}{|c|}{\textbf{B. Disturbances originated within the network}}\\ \hline
3   & 10 & \makecell[l]{whole\\ network} & \makecell[l]{14 trains \\ KS + IC}  & \makecell[l]{Arbitrary delays of 14 trains \\ delays up to 35 minutes} \\ \hline
\multicolumn{5}{|c|}{\textbf{C. Disturbances within the network and closures}}\\ \hline
4   & 11 &  KTC-CB & \makecell[l]{10 trains \\ KS + IC}   & \makecell[l]{double-track line KTC-CB closed \\ trains rerouted to  single-track line}\\ \hline
5   & 11 &  \makecell[l]{whole\\ network} & \makecell[l]{most\\ trains}  & \makecell[l]{double-track line KTC-CB closed as in 4 \\ + delays as in 3}\\ \hline
6  & 10 & \makecell[l]{whole\\ network} & all trains   & \makecell[l]{multiple-track KZ - KO(STM) and double-track \\ KO - KL changed into single-track + delays as in 3}\\ \hline
7   & 11 &  \makecell[l]{whole\\ network} & all trains   & \makecell[l]{multiple-track KZ - KO(STM) and double-track \\ KO - KL changed into single-track +  double-track \\ line KTC-CB closed as in 4 + delays as in 3}\\ \hline
8 & 11  & \makecell[l]{whole\\ network} & all trains   & \makecell[l]{Closures as in 7 + arbitrary delays of 13 trains \\ up to 30 minutes}   \\ \hline
9   & 11 & \makecell[l]{whole\\ network} & all trains   & \makecell[l]{Closures as in 7 + arbitrary delays of 15 trains \\ up to 30 minutes}\\ \hline
\end{tabular}
\caption{Real-life experiments. Networks of particular dispatching problems. In networks $0$ - $5$ most of the analyzed network is double-tracked, from network $6$ upward the number of single-track lines in the network increases at the expense of double and multiple-track lines. In each experiment instance, initial delays were chosen manually, as arbitrary values plausible in practice.  For each network, we considered a two-hour time with an afternoon rush-hour timetable from the year $2021$, with $\# J = 27$ trains. Besides the $\# S$ stations there are $2$ depots in each network.}
\label{tab::cases}
\end{table}

\begin{table}[tb]
	\centering
\begin{tabular}{|l|l|l|l|l|l|l|l|l|}
\hline
\multirow{3}{*}{n.} & \multicolumn{3}{l|}{CPLEX} & \multicolumn{3}{l|}{\begin{tabular}[c]{@{}l@{}} CQM hyb. {\tt t\_min} = $5$ s \\ mean value over $5$ realiz. \end{tabular}}   &  \multicolumn{2}{l|}{{\begin{tabular}[c]{@{}l@{}} comparison \\ CQM vs. CPLEX \end{tabular}}} \\ \hhline{~--------}
& \multicolumn{1}{l|}{\begin{tabular}[c]{@{}l@{}} \#vars / \\ \# constr.\end{tabular}} & \multicolumn{1}{l|}{obj.} & \begin{tabular}[c]{@{}l@{}} comp.\\ time [s]\end{tabular} &  \multicolumn{1}{l|}{obj.} & \begin{tabular}[c]{@{}l@{}} comp.\\ time [s]\end{tabular} & \multicolumn{1}{|l|}{\begin{tabular}[c]{@{}l@{}} QPU\\ time [s]\end{tabular}} & {\begin{tabular}[c]{@{}l@{}} obj. \\ diff. \% \end{tabular}} & {\begin{tabular}[c]{@{}l@{}} comp t. \\ diff \% \end{tabular}}\\ \hline
0& 556 /1756  & 0.0 & 0.07 & 0.0  & 5.19  & 0.03  & 0 \% & - \\ \hline
1& 556 / 1756  & 1.0 & 0.08 & 1.0  & 5.24  & 0.02  & 0 \% & - \\ \hline
2& 556 / 1740  & 6.0 & 0.08 & 6.0  & 5.37  & 0.03  & 0 \% & - \\ \hline
3& 556 / 1769  & 7.5   & 0.09 & 7.5  & 5.05  & 0.03  & 0 \% & - \\ \hline
4& 662 / 2210  & 78.25 & 0.20 & 82.70  & 5.10  &  0.03  & -5.6 \% & - \\ \hline
5& 662 / 2204  & 114.75& 0.25  & 132.55  & 5.25 &   0.02 & -15.5 \% & - \\ \hline
6& 711 / 2599  & 91.25 & 0.41  & 142.3  & 5.14 & 0.03  & -55.9 \% &  - \\ \hline
7& 817 / 3029  & 188.75 & 7.98  & 263.4  & 5.12 & 0.02  & -39.5 \% & 35.8 \% \\ \hline
8& 817 / 3074  & 157.75& 3.70  & 271.65  & 5.12 & 0.02 & -72.2 \% &  -38.4 \%  \\ \hline
9& 817 / 3081  & 185.5& 6.51  & 263.85  & 5.11 & 0.02  & -42.2 \% &  21.5 \%  \\ \hline
\end{tabular}
\caption{Real-life experiments. Results of ILP optimization on CPLEX and CQM D-Wave hybrid solver. $d_{\max} = 40$ was set for all networks. Computational times and QPU time were reported in D-Wave's output. We have selected {\tt t\_min} = $5$s as it provides the best balance between computational time and the quality of solution.  The last columns present a percentage comparison  (CQM - CPLEX ) / CPLEX, hence, the positive values represent CQM advantage, whereas negative values show CPLEX advantage. }\label{tab::results_of_CPLEX_CQM}
\end{table}

Table~\ref{tab::results_of_CPLEX_CQM} tabulates the results comparing CPLEX and CQM in terms of objective function value and computation time; problem sizes and QPU times are included. 
As CQM is probabilistic, $5$ runs have been performed for each network, yielding $5$ different realizations. This also facilitates the analysis of the statistics of the output and the differences between particular realizations.

The level of difficulty increases with the number of variables and constraints.
Observe that CQM always returns a feasible solution. 
For networks $0-3$ with initial delays only, CQM  generated optimal solutions, while in the remaining networks, the objective value was somewhat higher than the actual optima obtained with CPLEX. 
As for computational times, for  CPLEX they grow significantly - from $0.072$ to almost $8s$ (network $9$). Meanwhile, those of CQM remain almost constant at around $5.1s$. 
QPU times remain not greater than $0.032s$. 
Remarkably, for the hardest instances: networks $7$ and $9$, the CQM hybrid solver provides feasible solutions faster than exact CPLEX.
It does not find the actual optimum though, just a feasible solution close to it. 
Certainly, there can be other (meta)heuristics that provide solutions of similar quality in an even shorter time, for instance, CPLEX itself when running for a limited time, as we have found in the case of the synthetic examples in Section~\ref{sec::computation_generic}. (This experience is also iniline with cases of other similar scheduling problems, see the work of \cite{smierzchalski2024hybrid} for another example.) 
Nevertheless, the experience with these examples suggests that CQM provides useful results in reasonable time, so it is at least one of the possible heuristics worth considering.
We can expect benefits from the application of the hybrid solver on medium-scale railway networks with multiple closures,
as in networks 7-9. This finding follows our observation on synthetic data in Figure~\ref{fig::ko-glc-case1} b), where the benefits appeared for more complex instances, too. 

To evaluate the extent of difficulty of the problems analyzed in Table~\ref{tab::results_of_CPLEX_CQM},  Table~\ref{tab::stats} presents the values of the delays averaged over trains and realizations, resulting from CQM solutions, for selected stations (i.e. stations in the central part of the network in Figure~\ref{fig::network}). 
Delays with a broader impact are introduced starting from network $4$ onward, i.e. for networks with closures in a limited number of stations, see also Table~\ref{tab::cases}. 
From network $5$ onward, delays become more uniformly distributed among the stations when compared with results in Table~\ref{tab::cases}. 
However, for networks $7$ - $9$ the mean delays are the highest and almost uniformly distributed among all stations, showing that the spread of disruption impacts the whole network. 
These appeared to be more challenging network disruptions. 

\begin{figure}[ht]
\centering
\includegraphics[width = 0.22 \textwidth]{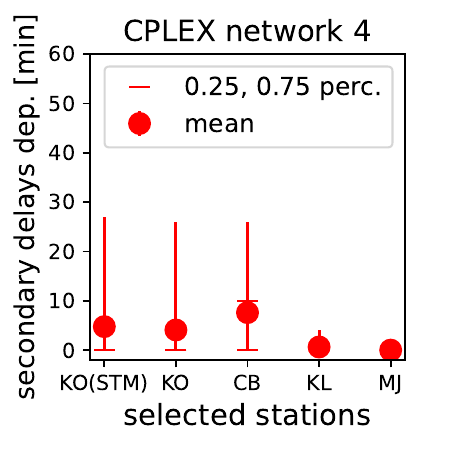}
\includegraphics[width = 0.22 \textwidth]{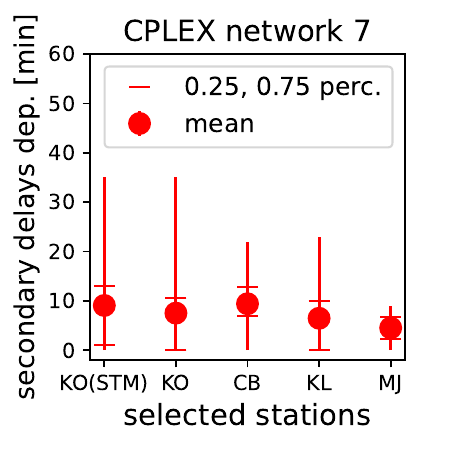}
\includegraphics[width = 0.22 \textwidth]{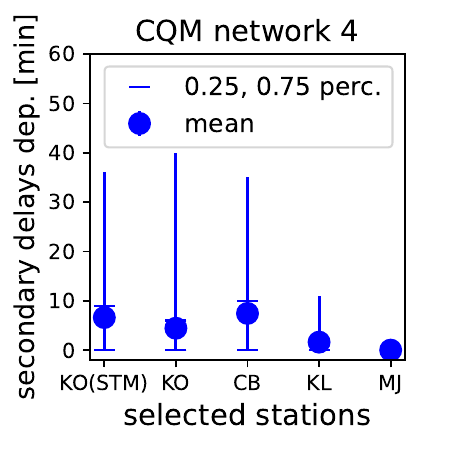}
\includegraphics[width = 0.22 \textwidth]{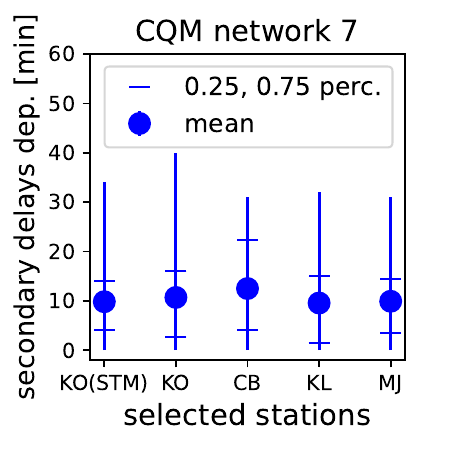}\caption{Real-life experiments. Statistics of CPLEX (left, read) and CQM (right, blue) solutions (Table~\ref{tab::results_of_CPLEX_CQM}) for selected networks in Table~\ref{tab::cases}. The statistics were calculated over trains in the classical cases, and over trains and realizations in the CQM cases. Vertical lines are ranges, while horizontal lines are $0.25$ and $0.75$ percentiles.}\label{fig::stats}
\end{figure}

Figure~\ref{fig::stats} shows a more detailed statistical analysis for two networks $4$ and $7$ and compares CPLEX (left, red) with exact CQM (right, blue). 
For network $4$, bigger delays occur mainly at stations KO(STM), KO, and CB for both CPLEX and CQM solutions.
This reflects the initial disturbances, which indeed occurred mainly between KO and CB (see Table~\ref{tab::cases}), and thus the spread is limited to these 3 stations. 
For network $7$, delays are more spread through the network due to disruption and initial disturbances. To conclude, we can expect an advantage from a hybrid (quantum-classical) approach, particularly in the case of more complex problems with extensive disruptions and disturbances occurring throughout the whole network. To facilitate further evaluation of the solutions, train diagrams are presented in Appendix A.
\begin{table}[tb]
\centering
\begin{tabular}{|l|l|l|l|l|l|}
\hline
 & \multicolumn{5}{c|}{ Delay [min]} \\ \hline
network & KO(STM) & KO   & CB   & KL   & MJ  \\ \hline
0    & 3.7     & 1.0  & 1.7  & 1.5  & 0.0 \\ \hline
1    & 3.1     & 1.3  & 2.1  & 1.5  & 0.0 \\ \hline
2    & 3.8     & 1.3  & 2.0  & 1.3  & 2.0 \\ \hline
3    & 2.6     & 2.1  & 1.4  & 0.7  & 1.5 \\ \hline
4    & 6.6     & 4.4  & 7.4  & 1.6  & 0.0 \\ \hline
5    & 7.1     & 6.2  & 10.7 & 1.0  & 2.7 \\ \hline
6    & 6.7     & 6.7  & 3.2  & 8.7  & 4.1 \\ \hline
7    & 9.8     & 10.7 & 12.5 & 9.5  & 9.9 \\ \hline
8    & 10.4    & 10.5 & 10.0 & 10.5 & 7.6 \\ \hline
9    & 10.0    & 10.3 & 11.5 & 9.8  & 9.6 \\ \hline
\end{tabular}
\caption{Real-life experiments. Average (over trains and realizations) secondary delay resulted from CQM solutions (Table~\ref{tab::results_of_CPLEX_CQM}) for all networks in Table~\ref{tab::cases}. Non-zero values for network $0$ come from counted delays in shunting movement in KO and KO(STM) that are not included in the objective, as well as some cases where the train gets a few minutes, delay but makes it up and leaves the network on time - these are typical issues for the railway traffic on the analyzed part of the network.}\label{tab::stats}
\end{table}

\begin{figure}
\centering
\includegraphics[width = 0.4\textwidth]{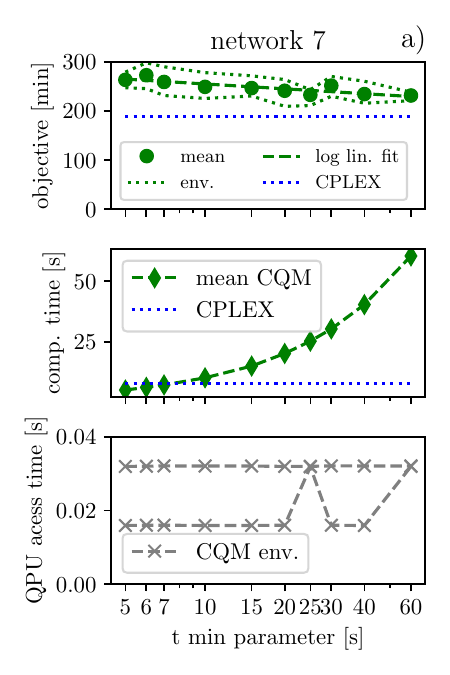}
\includegraphics[width = 0.4\textwidth]{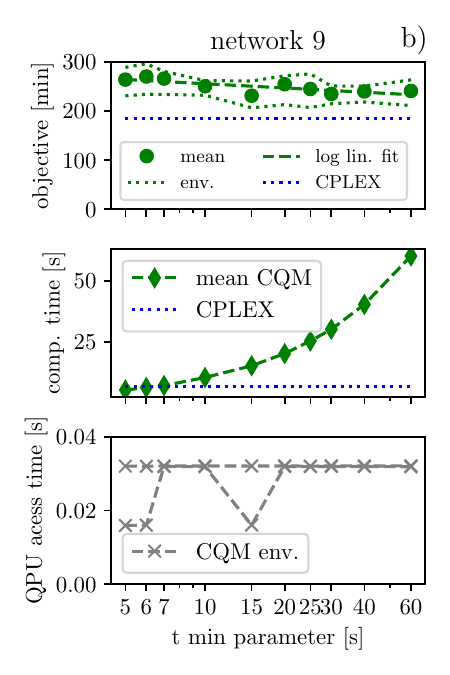}
\caption{Real-life experiments. Dependence on the {\tt t\_min}  parameter for network $7$ a) and network $9$ b), see Table~\ref{tab::cases}. All solutions were feasible. For higher {\tt t\_min} we expect smaller objective values but higher total computational time. For each parameter value, there were $5$ realizations of the experiment performed. Interestingly, a linear (negatively sloped) relation between the objective and the logarithm of ${\tt t\_min}$ can be observed, suggesting power law scaling.}\label{fig::sweep_tmin_case7}
\end{figure}

The right choice of the {\tt t\_min} parameter can improve results meaningfully. Let us analyze in more detail the role of the {\tt t\_min} solver parameter for selected problems.
In particular, for networks $7$ and $9$, we have performed runs with {\tt t\_min} sweeping over a range $[5,60]$; the results are presented in  Figure~\ref{fig::sweep_tmin_case7} a) and b). The objective value is roughly log-linear in computational time. For small {\tt t\_min} parameter values the computational time is shorter, whereas for large {\tt t\_min} it grows exponentially. Overall, the improvement in the objective is overwhelmed by the increase in computational time. This is important from the point of view of the algorithm layout, as it may be tied to a power law scaling. Such behavior is plausible in the case of QA methods~\citep{soriani2022three, albash2017temperature, domino2022statistical}. 

\section{Conclusions}
\label{sec:conclusions}

We have demonstrated that quantum annealers can be readily applicable in train rescheduling optimization in urban railway networks. In particular, we have encoded rescheduling problems in a linear integer program and solved these using a state-of-the-art solver (CPLEX), as well as the QA based CQM hybrid (quantum-classical) solver.
The results suggest that quantum computing and QA in particular, although an early-stage technology, are ready to tackle challenging railway rescheduling problems. 
While the CQM hybrid solver does not outperform classical solvers in general, we have found specific cases in which they were competitive. 
These results support the expectation that future quantum devices can become efficient in solving large-scale problems, like, in our case,  real-time railway dispatching problems on the national scale, even in the range that is beyond the scope of current exact models and heuristics.

While the quantum-based solvers possibly return suboptimal solutions, these are still feasible and often better than the solutions obtained manually or based on smaller-scale models. 
It is important to note that the CQM hybrid solver has always used the QPU in each computation. The actual assistance from the QPU may have boosted the classical heuristics in the hybrid solver.

A possible way of improving on the presented technique would be to find a way of determining the smallest suitable $d_{\text{max}}$ based on the input data instead of an ad-hoc choice, by using some fast enough methods. Apart from that, several future research directions can be determined. 
The first concerns the ongoing worldwide efforts on the development of fault-tolerant quantum computing devices \citep{divincenzo2000physical} with proper error correction/mitigation. 
Those future quantum devices are expected to
boost the efficiency of approaches supported by quantum computing (and QA in particular). 
The second concerns the creation of a custom open-source hybrid solver that is dedicated to railway problems and will use the noise of the quantum device to our advantage e.g. by modeling some stochastic behavior on railways. 
The third concerns the evolution and application of quantum-inspired techniques such as tensor network techniques or simulated bifurcation~\citep{goto2019quantum}. 
With all these future developments, QA-based approaches tend to become more powerful and ready to address for the first time large-scale real-time challenges in railway traffic management, which are currently unsolvable with traditional techniques.  

\section*{Data availability}

The code and the data used for generating the numerical results can be found in \url{https://github.com/iitis/railways_dispatching_silesia} under DOI  10.5281/zenodo.10657323

\section*{Acknowledgments} 

This research was supported by the Ministry of Culture and Innovation and the National Research, Development and Innovation Office within the Quantum Information National Laboratory of Hungary (Grant No. 2022-2.1.1-NL-2022-00004) (MK),  and by the Silesian University of Technology Rector’s Grant no. BKM - 720/RT2/2023     12/020/BKM\_2023/0252 (KK).
This research was funded in part by National Science Center, Poland, under grant number 2019/33/B/ST6/02011 (LB) and 2023/07/X/ST6/00396 (KD). For the purpose of Open Access, the authors have applied a CC-BY public copyright license to any Author Accepted Manuscript (AAM) version arising from this submission.

We would like to thank Sebastian Deffner as well as the whole scientific community of the Department of Physics, University of Maryland, Baltimore County (UMBC) for valuable discussion; and to Bartłomiej Gardas, and Zbigniew Puchała for valuable motivation; and Katarzyna Gawlak, Akash Kundu, and \"Ozlem Salehi for data validation and assistance with coding.
We acknowledge the cooperation with Koleje Ślaskie sp. z o.o. (eng. Silesian Railways).


\appendix

\section*{Appendix A.  Train diagrams of CPLEX and CQM}

\begin{figure}[tb]
\includegraphics[width = 0.32\textwidth]{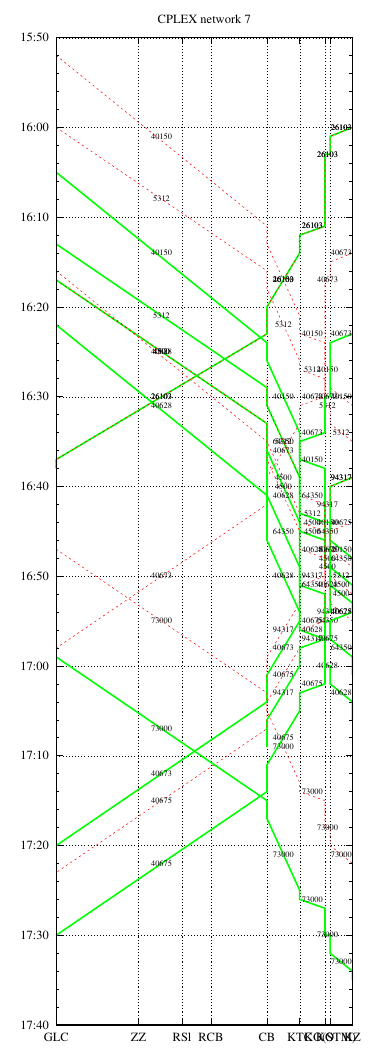}
\includegraphics[width = 0.32\textwidth]{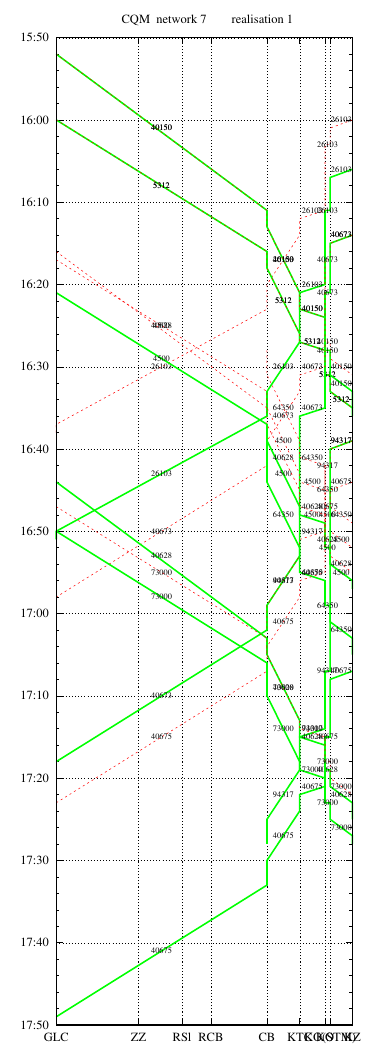}
\includegraphics[width = 0.32\textwidth]{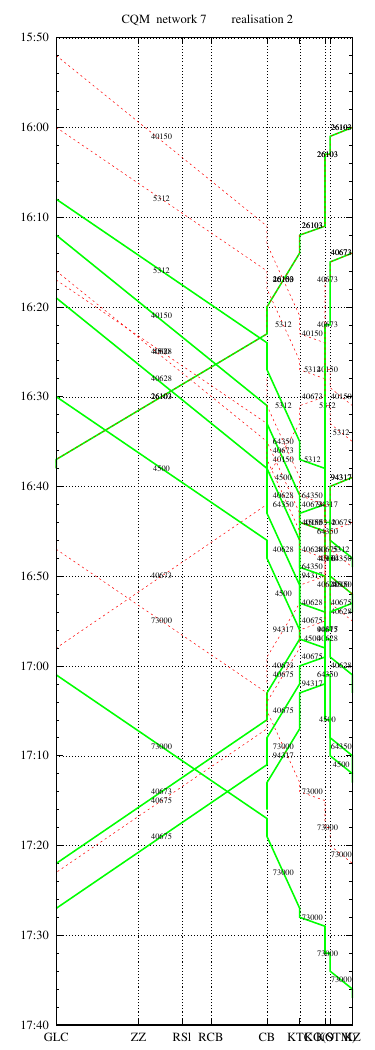}
\caption{Real-life experiment. Time-distance diagram solution of network $7$ CPLEX (left) and two realizations of CQM (middle and right), see Table~\ref{tab::cases}. Real-life experiment. Time-distance diagram CQM solution of network $7$ (see Table~\ref{tab::cases}) first realization. Closures (yielding single-track line) are between CB and KTC as well as between KO(STM) and KZ.}\label{fig::train_plot_ILP}
\end{figure}

To further evaluate the solutions, we present time-distance diagrams for network $7$, for the CPLEX optimal solution and two different quantum realizations; these can be seen in Figure~\ref{fig::train_plot_ILP} (left: exact CPLEX, middle and right: two realizations from CQM). We have chosen the line segment between GLC and KZ,  offering a good demonstration. 
(We have introduced a small artificial 'distance' between KO and KO(STM) which are contiguous locations for the sake of better visibility, hence the horizontal lines in the diagram.)
The red dotted lines represent the conflicted situation, and the green solid lines represent the solution. 
All three proposed solutions have similar figures of merit, they all provide a valid option to be realized. 
The conflicts are resolved in somewhat different ways, see e.g. the paths of trains 40150, 5312, and 26103. 
In details:
\begin{itemize}
    \item in the CPLEX solution, train 26103 passes first the KTC-CB closure heading for GLC, then 40150 followed by 5312 go in opposite directions,
    \item in the CQM solution, first realization, 40150 and 5312 pass first the KTC-CB closure and 26103 waits at KTC for them to pass,
    \item in the CQM solution, second realization, 26103 passes first KTC-CB closure, then 5312 followed by 40150 go in opposite directions.
\end{itemize}
This results in the following objective values: for CPLEX - $188.75$, for CQM first realization - $279$ for CQM second realization - $261.5$. Although the first solution is optimal, it is reasonable to supply dispatchers with the range of feasible solutions.
These diagrams illustrate that the CQM hybrid solver can produce usable solutions for railway traffic management.  
An additional benefit of the probabilistic nature of the quantum-based method is that it produces different alternatives after each sampling, without any modification. These could be offered to the dispatcher in a decision support system as possibilities. The dispatcher's choice can be influenced by other circumstances not covered by the present model.

\end{document}